\documentclass[aps,prd,twocolumn,superscriptaddress,nofootinbib,10pt]{revtex4-1}   

\pdfoutput=1
\usepackage[utf8]{inputenc}
\usepackage{amssymb}
\usepackage{amsmath}
\usepackage{amsfonts}
\usepackage{graphicx}
\usepackage{color}
\usepackage{xspace}
\usepackage{comment}
\usepackage{hyperref}
\usepackage[normalem]{ulem}
\usepackage{gensymb}
\usepackage{mathtools}
\usepackage[section]{placeins}
\usepackage{afterpage}
\usepackage{braket}
\usepackage{float}
\usepackage{slashed}
\usepackage{ulem}
\usepackage{appendix}

\usepackage{mathtools}
\newlength{\LHSwd}
\usepackage{tikz}
\usetikzlibrary{arrows.meta, decorations.pathmorphing}

\usepackage{cancel}

\usepackage{multirow,rotating}
\usepackage[dvipsnames]{xcolor}
\usepackage{orcidlink}   

\begin{document}
	
      \title{Entanglement measures and Bell-type spin-correlation observables in tau-lepton pairs at the Super Tau-Charm Facility}

      \author{Beizhi Yang\,\orcidlink{0009-0008-1507-3807}}
      \affiliation{School of Physics, Hefei University of Technology, Hefei 230601, People’s Republic of China}

      \author{Yu Zhang\,\orcidlink{0000-0001-9415-8252}}
      \email{dayu@hfut.edu.cn}
      \affiliation{School of Physics, Hefei University of Technology, Hefei 230601, People’s Republic of China}

      \author{Zeren Simon Wang\,\orcidlink{0000-0002-1483-6314}}
      \email{wzs@hfut.edu.cn}
      \affiliation{School of Physics, Hefei University of Technology, Hefei 230601, People’s Republic of China}

      \author{Xiaorong Zhou\,\orcidlink{0000-0002-7671-7644}}
      \affiliation{University of Science and Technology of China, Hefei 230026, People’s Republic of China}
      \affiliation{State Key Laboratory of Particle Detection and Electronics, Beijing 100049, Hefei 230026, People’s Republic of China}

	\begin{abstract}       
        Within the framework of quantum field theory and the Standard Model (SM), we investigate the prospects of studying entanglement measures and Bell-type spin-correlation observables in the electroweak process $e^- e^+\to \tau^-\tau^+$ at the center-of-mass (COM) energies of $\sqrt{s}=3.670,4.630$, and $7.000$~GeV at the proposed Super Tau-Charm Facility (STCF) in China. Focusing on the hadronic decay channel $\tau^\pm\to \pi^\pm \nu$, we determine the spin-correlation coefficients of the $\tau^-\tau^+$ system within the SM framework using measurable production kinematics, namely the $\tau$ velocities and scattering angles in the COM frame. From these correlation coefficients, we construct concurrence and a Bell-type correlation combination as defined in the literature. Assuming an integrated luminosity of 1~ab$^{-1}$ for each considered COM energy, and incorporating realistic detector efficiencies as well as statistical and systematic uncertainties, we estimate the expected sensitivity to these observables at STCF. Our results indicate that, under the SM hypothesis, the corresponding Bell-type correlation combinations could be resolved with high statistical significance at STCF.
 	\end{abstract}


    
	\maketitle
    \noindent
    
\section{Introduction}\label{sec:intro}

Quantum entanglement and Bell-type correlations are characteristic features of quantum theory; see Refs.~\cite{Brunner:2013est,Horodecki:2009zz} for reviews.
Their experimental investigation provides insight into quantum correlation structures and has traditionally proceeded through the study of Bell-type inequalities~\cite{Bell:1964kc,Clauser:1974tg,Aspect:1981zz};
In particular, Bohm and Aharonov~\cite{Bohm:1957zz} first presented the spin-spin realization of the Einstein-Podolsky-Rosen paradox~\cite{Einstein:1935rr}.
Over the past decades, numerous experiments, primarily at low energies using photons and atomic systems, have successfully explored such correlations; see, e.g., Refs.~\cite{Freedman:1972zza,Aspect:1981nv,Aspect:1982fx,Weihs:1998gy,Hagley:1997bob,Yin:2017ips,BIGBellTest:2018ebd}, as well as the review articles~\cite{Clauser:1978ng,Genovese:2005nw}.

In addition to low-energy experiments, the possibility of exploring entanglement measures and Bell-type spin-correlation combinations at high-energy particle colliders has attracted increasing attention. Various theoretical investigations and discussions have addressed this question; see, for example, Refs.~\cite{Tornqvist:1980af,Abel:1992kz,Benatti:1997fr,Benatti:1999jt,Benatti:1999du,Bertlmann:2001ea,Banerjee:2014vga,Acin:2000cs,Li:2008dk,Baranov:2008zzb,Chen:2013epa,Qian:2020ini,Banerjee:2015mha,Yongram:2013soa,Cervera-Lierta:2017tdt}; see also Ref.~\cite{Aspect2002} for a review.

In recent years, owing to advances in accelerator and detector technologies, increasingly large datasets have become available, leading to renewed interest in quantum-correlation studies at high-energy colliders.
In $B$-meson decays, analyses of entanglement measures and Bell-type correlation combinations have been reported~\cite{Fabbrichesi:2023idl}, followed by studies of quantum correlations in top-quark pair systems at the LHC~\cite{Afik:2020onf,ATLAS:2023fsd,CMS:2024pts}.
It has also been discussed whether Bell-type correlation combinations could be accessed in such systems~\cite{Fabbrichesi:2021npl}.
Numerous phenomenological works have further explored quantum correlations in elementary particle systems; see, e.g., Refs.~\cite{Aoude:2022imd,Afik:2022dgh,Fabbrichesi:2022ovb,Ehataht:2023zzt,Barr:2021zcp,Barr:2022wyq,Aguilar-Saavedra:2022wam,Fabbrichesi:2023cev,Severi:2021cnj,Larkoski:2022lmv,Aguilar-Saavedra:2022uye,Afik:2022kwm,Gong:2021bcp,Aguilar-Saavedra:2023hss,Aguilar-Saavedra:2024fig,White:2024nuc,Han:2024ugl,Fabbrichesi:2025ywl,Ashby-Pickering:2022umy,Aguilar-Saavedra:2022mpg,Altakach:2022ywa,Aoude:2023hxv,Morales:2023gow,Bernal:2023ruk,Bi:2023uop,Dong:2023xiw,Ma:2023yvd,Sakurai:2023nsc,Bernal:2023jba,Han:2023fci,Cheng:2023qmz,Aguilar-Saavedra:2024hwd,Aguilar-Saavedra:2024vpd,Aguilar-Saavedra:2024whi,Duch:2024pwm,Morales:2024jhj,Subba:2024mnl,Maltoni:2024csn,Afik:2025grr,Wu:2024asu,Cheng:2024btk,Gabrielli:2024kbz,Ruzi:2024cbt,Cheng:2024rxi,Wu:2024ovc,Ruzi:2024iqu,Altomonte:2024upf,Fabbrichesi:2024rec,Cheng:2025cuv,Han:2025ewp,Guo:2026yhz,Han:2023fci,Bernal:2024xhm,DelGratta:2025qyp,Goncalves:2025mvl,Ruzi:2025jql,Hong:2025drg,Goncalves:2025xer,Altakach:2022ywa,LoChiatto:2024dmx,Ruzi:2024cbt,Ding:2025mzj,Pei:2025yvr,Pei:2025ito,Cao:2025qua,Cheng:2025zcf,Shi:2016bvo,Shi:2019mlf,Shi:2019kjf,Pei:2026rlh,Pei:2026wfu,Pei:2026khg,Antozzi:2026vdi}.
See also Ref.~\cite{Barr:2024djo} for a recent review.
These studies extend beyond top-quark pairs to systems including pairs of tau leptons, hyperons, baryons, and gauge bosons from Higgs decays at various colliders.
Furthermore, decays of flavor states have been proposed as laboratories for investigating entanglement~\cite{Bertlmann:2001ea,Go:2003tx}.

To study entanglement measures and Bell-type spin-correlation combinations at colliders, suitable observables must be constructed, typically expressed in terms of spin-correlation coefficients.
In practice, many collider analyses infer such correlation information from the momenta and angular distributions of the decay products of the produced particles.

In this work, we follow a kinematic-reconstruction method proposed in Ref.~\cite{Cheng:2024rxi}, in which the relevant spin-correlation coefficients are computed from the velocity and polar angle of the produced particles, without explicit use of the decay-product angular distributions.
Compared with the common decay-based approaches, this kinematic method can potentially reduce statistical uncertainties~\cite{Han:2025ewp}.

In this work, we focus on the Super Tau-Charm Facility (STCF)~\cite{Achasov:2023gey,Ai:2025xop}, a next-generation $e^+e^-$ collider proposed to be constructed in Hefei, China, as a successor to the Beijing Spectrometer III (BESIII) experiment~\cite{BESIII:2009fln} at the Beijing Electron-Positron Collider.
The facility is designed to deliver GeV-scale electron and positron beams in a circular tunnel, enabling head-on collisions at center-of-mass (COM) energies ranging from $2$~GeV to $7$~GeV, with a peak luminosity of $0.5\times10^{35}$~cm$^{-2}$\,s$^{-1}$.
As its name suggests, the program is expected to produce large samples of Standard Model (SM) tau leptons and charm hadrons, including both mesons and baryons.

We will focus on studying entanglement measures and specific Bell-type spin-correlation observables within the SM framework in the electroweak process $e^- e^+\to \tau^-\tau^+$ at STCF.
Similar analyses involving the $\tau^- \tau^+$ final state have been proposed for LEP~\cite{Privitera:1991nz},\footnote{See also related discussions in Refs.~\cite{Abel:1992kz,Dreiner:1992gt}.} BESIII~\cite{Han:2025ewp}, and Belle~II~\cite{Ehataht:2023zzt} with $e^- e^+$ collisions, the LHC with $pp$ collisions~\cite{Fabbrichesi:2022ovb,Zhang:2025mmm}, Pb-Pb ultraperipheral collision~\cite{Lu:2025hwy}, and future lepton colliders~\cite{Altakach:2022ywa,Ma:2023yvd,Fabbrichesi:2024wcd}.

Before proceeding, we note that the interpretation of collider-based studies of entanglement measures and Bell-type spin-correlation combinations has been the subject of ongoing discussion in the literature (see, e.g., Refs.~\cite{Abel:1992kz,Bechtle:2025ugc,Abel:2025skj,Li:2024luk,Low:2025aqq}), particularly regarding their implications for foundational tests of locality and the exclusion of Local Hidden-Variable Theories (LHVTs).\footnote{See Appendix~A of Ref.~\cite{Bechtle:2025ugc} for a concise review of LHVTs.}
In particular, it has been shown that collider observables of this type can always be described by an explicit Local Hidden-Variable Theory (LHVT) construction, implying that Bell-type inequalities evaluated in such setups do not constitute genuine tests of locality~\cite{Abel:1992kz,Bechtle:2025ugc,Abel:2025skj,Li:2024luk,Low:2025aqq}.

In this paper, we therefore work within the standard quantum-field-theory (QFT) framework and focus on operationally defined observables constructed from spin-correlation coefficients predicted within the SM framework.
We emphasize that our analysis does not aim at a foundational test of locality; rather, it examines specific spin-correlation combinations within the SM and explores the experimental prospects for accessing entanglement measures and Bell-type spin-correlation observables at the proposed STCF, assessing their potential within its physics program.

Beyond their conceptual connection to entanglement studies, the spin-correlation coefficients considered here provide a direct probe of the spin structure predicted by QFT in a clean, quantum-electrodynamics-dominated regime.
At the STCF energies, the process $e^+e^- \to \tau^+\tau^-$ is overwhelmingly governed by photon exchange, leading to precise and theoretically well-controlled predictions for the correlation matrix within the SM.
Determining these coefficients experimentally therefore constitutes a direct validation of the QFT prediction for spin correlations in an elementary two-fermion system.
In addition, such measurements establish a well-controlled benchmark for collider-based entanglement observables, complementary to studies in hadronic systems or high-energy regimes, and could in principle probe deviations induced by higher-dimensional contact interactions or anomalous $\tau$ dipole couplings.

The remainder of this work is organized as follows.
In Sec.~\ref{sec:theo}, we present the theoretical framework for the entanglement measures and Bell-type spin-correlation observables considered here, and outline the kinematic method used in our analysis. 
In Sec.~\ref{sec:exp}, we briefly describe the STCF setup and explain the procedure used to evaluate the signal significances. 
Our numerical results are presented and discussed in Sec.~\ref{sec:results}. 
We summarize and conclude in Sec.~\ref{sec:conclusions}.

\section{Theoretical framework}\label{sec:theo}
\subsection{Entanglement measures and Bell-type spin-correlation observables}

The $\tau^-\tau^+$ pair produced in the signal process forms a two-qubit system, which can be described by a spin density matrix $\rho$.
It can be decomposed in the Fano-Bloch representation as follows~\cite{Fano:1983zz}:
\begin{eqnarray}
\rho&=&\frac{1}{4}\Big(\mathbb{I}_2\otimes\mathbb{I}_2+\sum_i{B_i^+\sigma_i\otimes\mathbb{I}_2}\nonumber\\
&&+\sum_j{B_j^-\mathbb{I}_2\otimes\sigma_j}+\sum_{ij}{C_{ij}\sigma_i\otimes\sigma_j}\Big),
\end{eqnarray}
where $i,j=1,2,3$.
The Fano coefficients $B_i^+$ and $B_j^-$ encode the net polarization of the first and the second qubits, respectively, and $C_{ij}$ denotes the spin-correlation matrix. $\otimes$ denotes a tensor product and $\sigma_i$ are the Pauli matrices.

To characterize the spin-correlation structure of the final state, one needs to determine the 15 coefficients $B_i^+,B_j^-$, and $C_{ij}$.
Requiring CP invariance implies $B^+ = B^-$ and $C = C^T$.
In this work, we neglect single-$\tau$ polarization effects and set $B^+=B^-=0$.
At the COM energies of the electron-positron collisions at STCF, the process $e^-e^+\to\tau^-\tau^+$ is overwhelmingly dominated by photon exchange, while $Z$-boson exchange and $\gamma$--$Z$ interference effects are suppressed by a factor of $s/m_Z^2 \sim \mathcal{O}(10^{-4}\text{--}10^{-3})$ with $m_Z$ denoting the $Z$-boson mass.
The resulting single-$\tau$ polarization is therefore negligible for the present analysis.
For the unpolarized process $e^- e^+\to \tau^- \tau^+$, only four independent components of the spin-correlation matrix $C_{ij}$ remain: $C_{11}, C_{22}, C_{33}$ and $C_{13}$, with $C_{31}=C_{13}$; all other components vanish~\cite{Maltoni:2024csn}.
See also Sec.~\ref{subsec:analysis_Cij} for further detail.

To characterize entanglement measures and Bell-type spin-correlation combinations, we define the concurrence $\mathcal{C}$ and a Bell-type correlation variable $\mathcal{B}$ as
\begin{eqnarray}
    \mathcal{C}&=&\frac{1}{2}(C_{11}+C_{22}+C_{33}-1),\\
    \mathcal{B}&=&\sqrt{2}(C_{33}-C_{22}).
\end{eqnarray}

The concurrence satisfies $\mathcal{C}\in[0,1]$, with $\mathcal{C}>0$ corresponding to a non-separable state within the quantum-mechanical description.
For the Bell-type correlation variable, values of $\mathcal{B} > 2$ correspond, within the SM spin-correlation framework assumed here, to combinations that would exceed the classical Clauser-Horne-Shimony-Holt bound if interpreted in a two-qubit setting.

\subsection{Analysis of the spin correlation matrix \\in $e^- e^+\to \tau^-\tau^+$}\label{subsec:analysis_Cij}

To analyze the spin-correlation structure, we should choose a basis in which to quantize the spin.
Several choices of spin-quantization bases are commonly used, including the beam basis~$\{\hat{x},\hat{y},\hat{z}\}$, the helicity basis~$\{\hat{k},\hat{r},\hat{n}\}$, and the diagonal basis~$\{\hat{e}_1,\hat{e}_2,\hat{e}_3\}$~\cite{Cheng:2023qmz,Cheng:2024btk}.
In the following, we work with the diagonal beam basis, as this choice maximizes the spin-correlations coefficients.
\begin{figure}[t]
    \centering
    \includegraphics[width=0.495\textwidth]{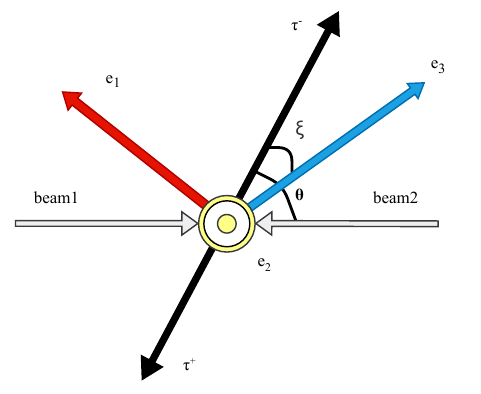}
    \caption{The $\tau$-pair system in the diagonal beam basis which can be converted to the helicity basis by a rotation of the angle $\xi$.}
    \label{fig:dia}
\end{figure}
To define the diagonal basis, we start from the helicity basis which is defined relative to the $\tau^-$ momentum direction $\hat{k}$, and then rotate the basis by an event-dependent angle $\xi$ in the scattering plane.
We illustrate the $\tau^- \tau^+$-pair system in the diagonal beam basis in Fig.~\ref{fig:dia}.
Here, $\tan{\xi}=\sqrt{1-\beta^2}\tan\theta$, with $\beta$ and $\theta$ denoting the speed and scattering angle of the $\tau$-lepton in the COM frame relative to the incoming electron beam.

Further, in the diagonal basis, we may consider the mixed state
\begin{eqnarray}
    \rho_{\text{mixed}}&=&\frac{1}{2}\ket{\uparrow\uparrow}\bra{\uparrow\uparrow}+\frac{1}{2}\ket{\downarrow\downarrow}\bra{\downarrow\downarrow}\,,
\end{eqnarray}
and the pure Bell state
\begin{eqnarray}
    \rho_{\text{pure}}&=&\ket{\psi^+}\bra{\psi^+},\text{ with }\ket{\psi^+}=\frac{1}{\sqrt{2}}(\ket{\uparrow\uparrow}+\ket{\downarrow\downarrow})\,,
\end{eqnarray}
where $\uparrow$ ($\downarrow$) represents the positive (negative) eigenvalue of the $\hat{e}_3$ direction in the diagonal basis.
The density matrix relevant for the process can be expressed as a linear combination of these two states:
\begin{eqnarray}
    \rho = (1-\lambda)\rho_{\text{mixed}}+\lambda\, \rho_{\text{pure}}, \,\,\,\, \lambda = \frac{\beta^2 \sin^2{\theta}}{2-\beta^2\sin^2{\theta}}. \label{eqn:rho_decomposition}
\end{eqnarray}
We stress that Eq.~\eqref{eqn:rho_decomposition} refers to the production-level spin density matrix defined in the $\tau^+\tau^-$ COM frame.
In the present analysis, the quantities $\beta$ and $\theta$ are reconstructed from measurable kinematic variables in that frame using the kinematic reconstruction method described here; consequently, no explicit boost of the pion momenta into the individual $\tau$ rest frames is required for the construction of the spin-correlation coefficients.

For a purely $s$-channel process, the spin-correlation structure is determined by the underlying interaction tensor.
Following Ref.~\cite{Maltoni:2024csn}, we decompose the spin-correlation matrix elements into three contributions,
\begin{equation}
    C_{ij} = C_{ij}^{[0]} + C_{ij}^{[1]} + C_{ij}^{[2]},
\end{equation}
where $i,j=1,2,3$ label matrix-element indices.

\begin{figure}[t]
    \centering
    \includegraphics[width=0.495\textwidth]{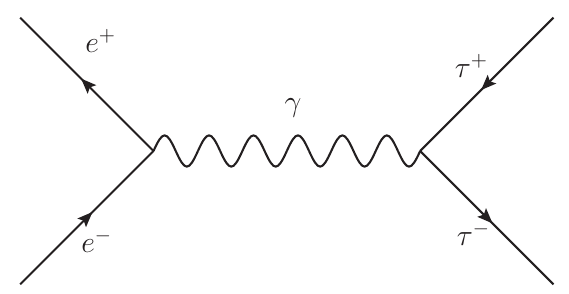}
    \caption{Leading tree-level Feynman diagram for the $e^- e^+ \to \tau^- \tau^+$ process.
Higgs- and $Z$-boson-mediated contributions are negligible at the STCF COM energies and are therefore omitted.}
    \label{fig:feynman}
\end{figure}

The leading tree-level Feynman diagram for the process $e^+e^- \to \tau^+\tau^-$ is shown in Fig.~\ref{fig:feynman}, where the interaction proceeds via an $s$-channel SM photon.
In addition, Higgs- and $Z$-boson-mediated diagrams are present in principle; however, their contributions are negligible at the relatively low STCF COM energies ($\sqrt{s}\ll m_Z, m_h$), and the Higgs exchange is further suppressed by the small electron Yukawa coupling.
These contributions are therefore omitted in the present analysis.

For symmetric electron–positron head-on collisions at STCF, the kinematics of the $\tau^-\tau^+$ final state can be fully characterized by two variables: the scattering angle $\theta$ in the COM frame (used to construct the diagonal basis) and the $\tau$-lepton speed $\beta$, given by
\begin{equation}
    \beta = \sqrt{1 - \frac{4\,m_{\tau}^2}{s}}.
\end{equation}

Introducing the shorthand notations $s_\theta \equiv \sin\theta$ and 
$c_\theta \equiv \cos\theta$, the spin-correlation coefficients can be written as~\cite{Maltoni:2024csn}
\begin{equation}
    C_{ij} = \frac{\widetilde{C}_{ij}}{A}=\frac{\widetilde{C}_{ij}^{[0]} + \widetilde{C}_{ij}^{[1]} + \widetilde{C}_{ij}^{[2]}}{A^{[0]} + A^{[1]} + A^{[2]}},
    \label{eqn:Cij}
\end{equation}
where
\begin{align}
&\left\{
\begin{aligned}
\mathmakebox[\LHSwd][l]{A^{[0]}} &=
F^{[0]}\!\left(\beta^2 c_{\theta }^2-\beta^2+2\right), \\
\mathmakebox[\LHSwd][l]{\widetilde{C}^{[0]}_{11}} &=
F^{[0]}\!\left[\beta^2-\left(\beta^2-2\right)c_{\theta }^2\right], \\
\mathmakebox[\LHSwd][l]{\widetilde{C}^{[0]}_{13}} &=
2\,F^{[0]}\sqrt{1-\beta^2}\,c_{\theta }\,s_{\theta }, \\
\mathmakebox[\LHSwd][l]{\widetilde{C}^{[0]}_{22}} &=
F^{[0]}\, s_{\theta }^2\, \beta^2, \\
\mathmakebox[\LHSwd][l]{\widetilde{C}^{[0]}_{33}} &=
-F^{[0]}(2-\beta^2)\,s_{\theta }^2,
\end{aligned}
\right.
\label{eqn:A0C0}
\\[0.1em]
&\left\{
\begin{aligned}
\mathmakebox[\LHSwd][l]{A^{[1]}} &=
2\,F^{[1]}\,c_{\theta}, \\
\mathmakebox[\LHSwd][l]{\widetilde{C}^{[1]}_{11}} &=
2\,F^{[1]}\,c_{\theta}, \\
\mathmakebox[\LHSwd][l]{\widetilde{C}^{[1]}_{13}} &=
F^{[1]}\sqrt{1-\beta^2}\,s_{\theta},\\
\mathmakebox[\LHSwd][l]{\widetilde{C}^{[1]}_{22}} &=
0, \\
\mathmakebox[\LHSwd][l]{\widetilde{C}^{[1]}_{33}} &=
0,
\end{aligned}
\right.
\label{eqn:A1C1}
\\[0.1em]
&\left\{
\begin{aligned}
\mathmakebox[\LHSwd][l]{A^{[2]}} &=
F^{[2]}\left(1+c_{\theta}^2\right), \\
\mathmakebox[\LHSwd][l]{\widetilde{C}^{[2]}_{11}} &=
F^{[2]}\left(1+c_{\theta}^2\right), \\
\mathmakebox[\LHSwd][l]{\widetilde{C}^{[2]}_{13}} &=
0, \\
\mathmakebox[\LHSwd][l]{\widetilde{C}^{[2]}_{22}} &=
F^{[2]}\, s_{\theta}^2, \\
\mathmakebox[\LHSwd][l]{\widetilde{C}^{[2]}_{33}} &=
-F^{[2]} \,s_{\theta}^2.
\end{aligned}
\right.
\label{eqn:A2C2}
\end{align}
We note that $C_{31}=C_{13}$, while all other components not listed above vanish.

In Eqs.~(\ref{eqn:A0C0}--\ref{eqn:A2C2}), the quantities $F^{[0]}$, $F^{[1]}$, and $F^{[2]}$ are defined as
\begin{eqnarray}
    F^{[0]} &=& 48 \Big( Q_{\tau}^2 Q_e^2 + 2 \,\text{Re} \,\frac{ 4 Q_{\tau} Q_e g_{V\tau} g_{Ve} m_{\tau}^2 }{c_W^2 s_W^2 (4 m_{\tau}^2-(1-\beta^2) m_Z^2 )} \nonumber\\
    &&+ \frac{16 g_{V\tau}^2 m_{\tau}^4 (g_{Ve}^2 + g_{Ae}^2)}{|c_W^2 s_W^2 (4 m_{\tau}^2-(1-\beta^2) m_Z^2 )|^2} \Big),\\
    F^{[1]} &=& 192 g_{A\tau} g_{Ae} m_{\tau}^2 \beta \Big(\frac{16 g_{V\tau} g_{Ve} m_{\tau}^2 }{|c_W^2 s_W^2 (4 m_{\tau}^2-(1-\beta^2) m_Z^2 )|^2} \nonumber\\
    &&+ 2 \text{Re} \frac{Q_\tau Q_e }{c_W^2 s_W^2 (4 m_{\tau}^2-(1-\beta^2) m_Z^2 )} \Big),\\
    F^{[2]} &=& \frac{768 g_{A\tau}^2 m_{\tau}^4 \beta^2 (g_{Ve}^2 + g_{Ae}^2)}{|c_W^2 s_W^2 (4 m_{\tau}^2-(1-\beta^2) m_Z^2 )|^2}.
\end{eqnarray}
Here, $Q_{\tau}=Q_e=-1$.
The parameters $c_W$ and $s_W$ denote the cosine and sine of the weak mixing angle, respectively, and $m_\tau$ and $m_Z$ are the masses of the $\tau$ lepton and the $Z$ boson.
The vector and axial-vector couplings of charged leptons to the $Z$-boson are denoted by $g_V$ and $g_A$, with
\begin{eqnarray}
    g_{V\tau}&=&g_{Ve}=\frac{I_3}{2}-Q_e s_W^2 \simeq -0.0188, \\
    g_{A\tau}&=&g_{Ae}=\frac{I_3}{2}=-\frac{1}{4},
\end{eqnarray}
where $I_3=-\frac{1}{2}$ is the weak isospin of the charged leptons.

The spin-correlation matrix for the unpolarized process $e^- e^+ \to \tau^- \tau^+$ then follows directly from the expressions given above.

\section{Experimental setup}\label{sec:exp}

At STCF, we consider three representative COM energies, $\sqrt{s}=3.670, 4.630$, and $7.000$~GeV, each with an integrated luminosity of 1~ab$^{-1}$.
These correspond to total $\tau^- \tau^+$ production yields of $N_{\text{tot}}=2.4\times 10^9, 3.4\times 10^9$, and $1.7\times 10^9$ events, respectively~\cite{Achasov:2023gey}.

At these COM energies, the contributions from $Z$- and Higgs-boson exchange are negligible. 
The differential cross section for $e^- e^+ \to \tau^- \tau^+$ is therefore well approximated by the photon-exchange contribution,
\begin{eqnarray}
    \frac{d\sigma}{d\cos\theta}=\frac{\pi\alpha^2\beta}{2s}(2-\beta^2\sin^2\theta).
    \label{eqn:diff_xs_ee2tautau_simplified}
\end{eqnarray}

\begin{table}[t]
\centering
\begin{tabular}{|c | c | c |} 
 \hline
 $\sqrt{s}$~[GeV]  & $\sigma(e^-e^+\to \tau^-\tau^+)$~[fb] & $N_S$ \\ 
 \hline
   3.670 & $2.4\times 10^6$ &  $2.4\times 10^9$\\
   4.630 & $3.4\times 10^6$ &  $3.4\times 10^9$\\
   7.000 & $1.7\times 10^6$ &  $1.7\times 10^9$\\
  \hline
\end{tabular}
\caption{Total cross sections for $e^- e^+\to\tau^- \tau^+$ at $\sqrt{s}=3.670$, $4.630$, and $7.000$~GeV, computed from Eq.~\eqref{eqn:diff_xs_ee2tautau_simplified}, together with the corresponding signal yields for an integrated luminosity of $1~\mathrm{ab}^{-1}$. The resulting event numbers are consistent with those reported in Ref.~\cite{Achasov:2023gey}.}
\label{table:XS_NS}
\end{table}
Integrating Eq.~\eqref{eqn:diff_xs_ee2tautau_simplified} over $\cos\theta$ yields the total cross sections shown in Table~\ref{table:XS_NS}, together with the corresponding signal yields for $1~\mathrm{ab}^{-1}$.

In the low-COM-energy limit relevant for STCF, the spin-correlation matrix in Eq.~\eqref{eqn:Cij} simplifies to~\cite{Ehataht:2023zzt}
\begin{eqnarray}
C_{ij}&&=\frac{1}{2-\beta^2\sin^2\theta}\\
&&\begin{pmatrix}
(2-\beta^2)\sin^2\theta & 0 & \sqrt{1-\beta^2}\sin(2\theta) \\
0 & -\beta^2\sin^2\theta & 0 \\
\sqrt{1-\beta^2}\sin(2\theta) & 0 & \beta^2
\end{pmatrix}.\nonumber
\end{eqnarray}
From this expression, the concurrence $\mathcal{C}$ and the Bell-type correlation variable $\mathcal{B}$ are obtained as
\begin{eqnarray}
    \mathcal{C}=\frac{\beta^2\sin^2\theta}{2-\beta^2\sin^2\theta}, \quad 
    \mathcal{B}=2\sqrt{1+\mathcal{C}^2}\, \,\,.  \label{eqn:CandB}
\end{eqnarray}

\begin{figure}[t]
    \centering
    \includegraphics[width=0.495\textwidth]{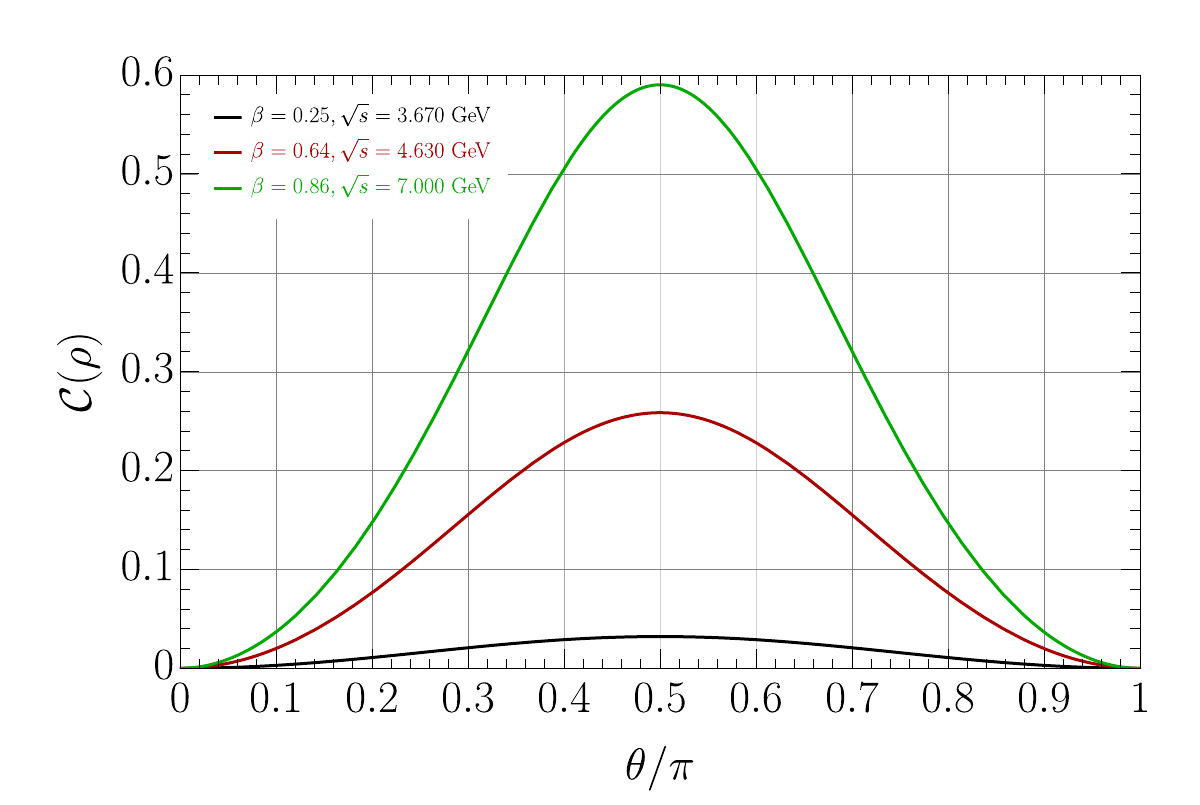}
    \includegraphics[width=0.495\textwidth]{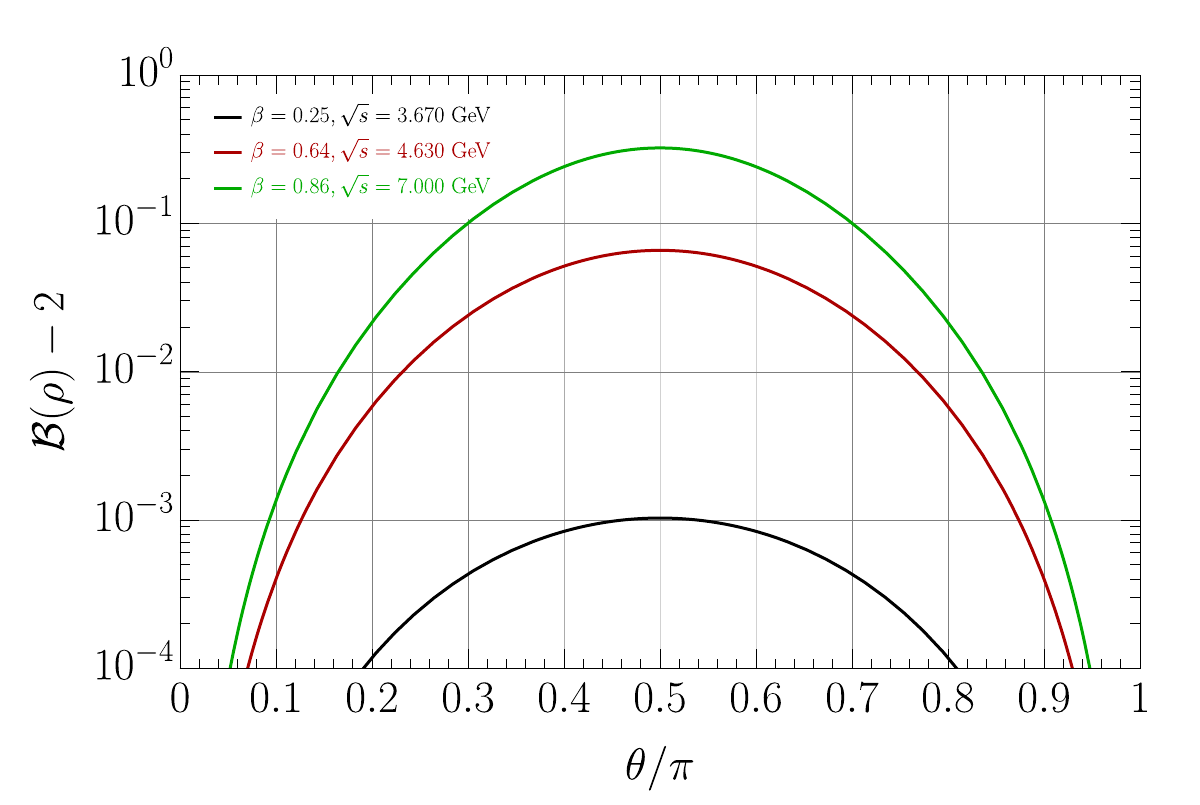}
    \caption{The concurrence $\mathcal{C}$ and the quantity $\mathcal{B}-2$ as functions of the scattering angle $\theta$ for COM energies of 3.670, 4.630, and 7.000 GeV.}
    \label{fig:B_C}
\end{figure}
Using Eq.~\eqref{eqn:CandB}, we show in Fig.~\ref{fig:B_C} the angular dependence of $\mathcal{C}$ and $\mathcal{B}-2$ for the considered COM energies.
For larger COM energies, the $\tau$-leptons are more boosted, corresponding to larger values of $\beta$, which leads to increased values of both $\mathcal{C}$ and $\mathcal{B}$.
Also, both observables reach their maximum at $\theta=\pi/2$.

Before proceeding with the computation, we briefly explain how the kinematics of the $\tau$-leptons can be reconstructed at STCF. 
In a clean $e^- e^+$ collider environment such as STCF, the $\tau$-leptons decay before they can be directly observed, and only their visible decay products (e.g.\ $\pi^\pm$) are detected. 
Since the four-momenta of the incoming electron and positron beams are known, energy–momentum conservation together with on-shell mass constraints allows the parent $\tau$-lepton velocity and scattering angle to be inferred from the measured pion momentum by solving the corresponding kinematic equations. 
This procedure generally leads to a two-fold algebraic ambiguity in the reconstructed $\tau$ momentum, which can be resolved using physical consistency criteria or treated statistically. 
Such constrained momentum-reconstruction techniques have been extensively developed for $\tau$ studies at lepton colliders (see, e.g., Refs.~\cite{Jeans:2015vaa,Bodrov:2024wrw,Kuhn:1993ra}).

To optimize the statistical sensitivity to the entanglement measures and Bell-type spin-correlation observables considered here, we impose phase-space cuts on the scattering angle $\theta$ around $\theta=\pi/2$.
Concretely, we restrict the angular range to
\begin{eqnarray}
    \frac{\pi}{2}-\frac{\Delta\theta}{2}<\theta <\frac{\pi}{2}+\frac{\Delta\theta}{2}.
    \label{eqn:theta_cut_window}
\end{eqnarray}

The corresponding cut efficiency factor is given by
\begin{eqnarray}
    R(\Delta\theta)=\frac{\int_{0}^{\sin(\frac{\Delta\theta}{2})}(2-\beta^2+\beta^2\cos^2\theta)\,d\cos\theta}{\int_0^1(2-\beta^2+\beta^2\cos^2\theta) \,d\cos\theta}\,\,.
\end{eqnarray}

We define the statistical significances associated with $\mathcal{C}$ and $\mathcal{B}$ as
\begin{eqnarray}
    \mathcal{S}(\mathcal{C})&\equiv&\frac{\overline{\mathcal{C}}}{\Delta\mathcal{C}_{\text{tot}}}=
    \frac{\overline{\mathcal{C}}}{\sqrt{\Delta\mathcal{C}_{\text{stat}}^2+\Delta\mathcal{C}_{\text{sys}}^2}},   \label{eqn:significanceC}\\
    \mathcal{S}(\mathcal{B})&\equiv&\frac{\overline{\mathcal{B}}}{\Delta\mathcal{B}_{\text{tot}}}=\frac{\overline{\mathcal{B}}-2}{\sqrt{\Delta\mathcal{B}_{\text{stat}}^2+\Delta\mathcal{B}_{\text{sys}}^2}}.  \label{eqn:significanceB}
\end{eqnarray}
Here, $\Delta\mathcal{C}_{\text{stat}}$ and $\Delta\mathcal{B}_{\text{stat}}$ denote the statistical uncertainties, while $\Delta\mathcal{C}_{\text{sys}}$ and $\Delta\mathcal{B}_{\text{sys}}$ represent the systematic uncertainties. 
We parameterize the relative systematic uncertainty as $\Delta_{\text{sys}}=\Delta C_{ij}/C_{ij}$~\cite{Cheng:2024rxi}.
\begin{eqnarray}
\Delta\mathcal{C}_{\text{stat}}&=&\frac{\sqrt{\text{Var}(\mathcal{C})}}{\sqrt{N}}\,, \label{eqn:DeltaCstat}\\
\Delta\mathcal{C}_{\text{sys}}&=&\frac{\Delta_{\text{sys}}}{2}\,\sqrt{\sum_{i} C_{ii}^2}\,, \label{eqn:DeltaCsys}\\
\Delta\mathcal{B}_{\text{stat}}&=&\frac{\sqrt{\text{Var}(\mathcal{B})}}{\sqrt{N}}\,, \label{eqn:DeltaBstat}\\
\Delta\mathcal{B}_{\text{sys}}&=&\sqrt{2}\,\Delta_{\text{sys}}\,\sqrt{C_{22}^2+C_{33}^2}\, \label{eqn:DeltaBsys},
\end{eqnarray}
where the variance is defined as
\begin{eqnarray}
\text{Var}(X)&=&\langle(\bar{X}-X)^2\rangle\nonumber\\
&=&\frac{1}{\Delta\theta}\int_{\frac{\pi}{2}-\frac{\Delta\theta}{2}}^{\frac{\pi}{2}+\frac{\Delta\theta}{2}}X^2(\theta)\,d\theta\nonumber\\
&&-\Big(\frac{1}{\Delta\theta}\int_{\frac{\pi}{2}-\frac{\Delta\theta}{2}}^{\frac{\pi}{2}+\frac{\Delta\theta}{2}}X(\theta)\,d\theta\Big)^2\,.
\end{eqnarray}
We proceed to calculate $N$ and the expectation values of $C_{ij}$, $\mathcal{C}$, and $\mathcal{B}$.
The expected number of signal events after the angular cut is estimated as
\begin{eqnarray}
    N(\Delta\theta)=\epsilon \cdot N_{\text{tot}} \cdot R(\Delta\theta)\cdot \Big(\text{BR}(\tau^- \to \nu_\tau \pi^-)\Big)^2.
\end{eqnarray}
Here, $\epsilon$ denotes the reconstruction efficiency of the final states.
Following the analyses in Ref.~\cite{Sun:2024vcd}, we take $\epsilon=0.3$ which includes the efficiencies of both tracking and a missing-energy cut to suppress hadronic backgrounds.
Also, $\text{BR}(\tau^- \to \nu_\tau \pi^-) \approx 0.108$~\cite{ParticleDataGroup:2024cfk}.\footnote{In principle, including additional $\tau$ decay channels such as $\tau^\pm\to \pi^\pm \pi^0 \nu$, $\tau^\pm\to \pi^\pm\pi^\pm\pi^\mp \nu$, and even $\tau^\pm\to e^\pm/\mu^\pm \nu\nu$, can enhance the signal rates by order~1. However, in this work, for simplicity we choose to confine ourselves to the $\tau^\pm\to \pi^\pm \nu$ channel.}
The squared branching ratio reflects the requirement that both the $\tau$-leptons decay to $\pi\nu_\tau$.

Using the kinematic method, the angularly averaged observables are computed as
\begin{eqnarray}
    \overline{\mathcal{C}}=\frac{1}{\Delta\theta}\int_{\frac{\pi}{2}-\frac{\Delta\theta}{2}}^{\frac{\pi}{2}+\frac{\Delta\theta}{2}}\mathcal{C}\,d\theta, \quad 
    \overline{\mathcal{B}}=\frac{1}{\Delta\theta}\int_{\frac{\pi}{2}-\frac{\Delta\theta}{2}}^{\frac{\pi}{2}+\frac{\Delta\theta}{2}}\mathcal{B}\,d\theta.
\end{eqnarray}
Similarly, the averaged spin-correlation coefficients are given by
\begin{equation}
    \overline{C_{ij}}=\frac{1}{\Delta\theta}\int_{\frac{\pi}{2}-\frac{\Delta\theta}{2}}^{\frac{\pi}{2}+\frac{\Delta\theta}{2}}C_{ij}\,d\theta.
\end{equation}

Before concluding this section, we briefly discuss potential background sources.
The dominant detector-related backgrounds at STCF arise from imperfect track reconstruction and particle identification, which can distort the measured pion momentum and consequently bias the inferred $\tau$-lepton kinematics and spin-correlation observables. 
Experience from existing $e^+ e^-$ collider experiments, in particular Belle~II~\cite{Ehataht:2023zzt,Belle-II:2023izd}, indicates that misidentification of charged tracks, together with finite momentum and angular resolution, can lead to incorrect reconstruction of the parent $\tau$ momentum.
Such reconstruction effects generally dominate over other detector-related backgrounds.

At the considered COM energies, physics backgrounds such as the $e^+e^-\to q\bar q$ continuum (including open-charm production at $\sqrt{s}=4.630$ and $7.000$~GeV) are kinematically allowed. 
However, for the simple one-prong topology $\tau^\pm\to\pi^\pm\nu_\tau$ considered here, these backgrounds can be efficiently suppressed through event-topology selections and particle-identification criteria. 
A detailed quantitative evaluation of these effects will ultimately require a dedicated full detector simulation, which lies beyond the scope of the present work.

\section{Results}\label{sec:results}

To evaluate the statistical significances $\mathcal{S}(\mathcal{C})$ and $\mathcal{S}(\mathcal{B})$, we consider benchmark values of the relative systematic uncertainty,
\begin{equation}
    \Delta_{\text{sys}}=\{0.5\%,1\%,2\%,5\%\},
\end{equation}
and use Eqs.~\eqref{eqn:significanceC} and~\eqref{eqn:significanceB}.

\begin{figure}[t]
    \centering
    \includegraphics[width=0.495\textwidth]{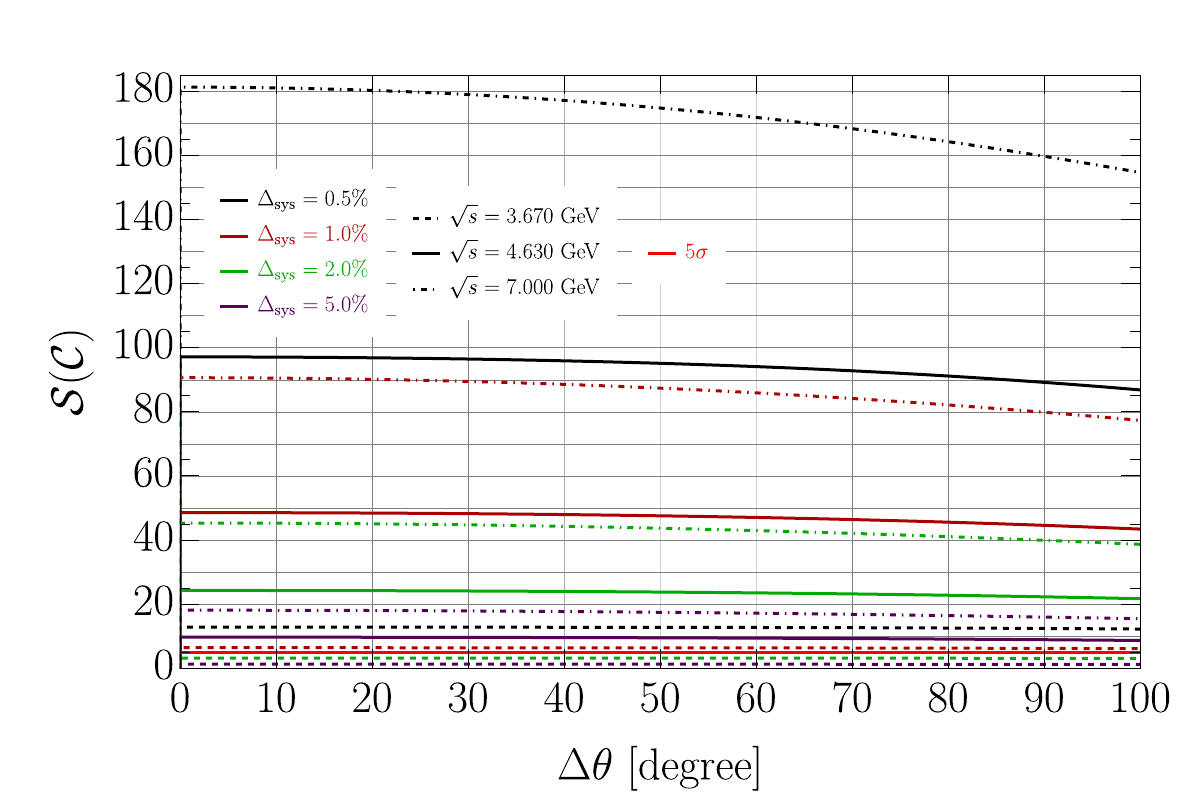}
    \includegraphics[width=0.495\textwidth]{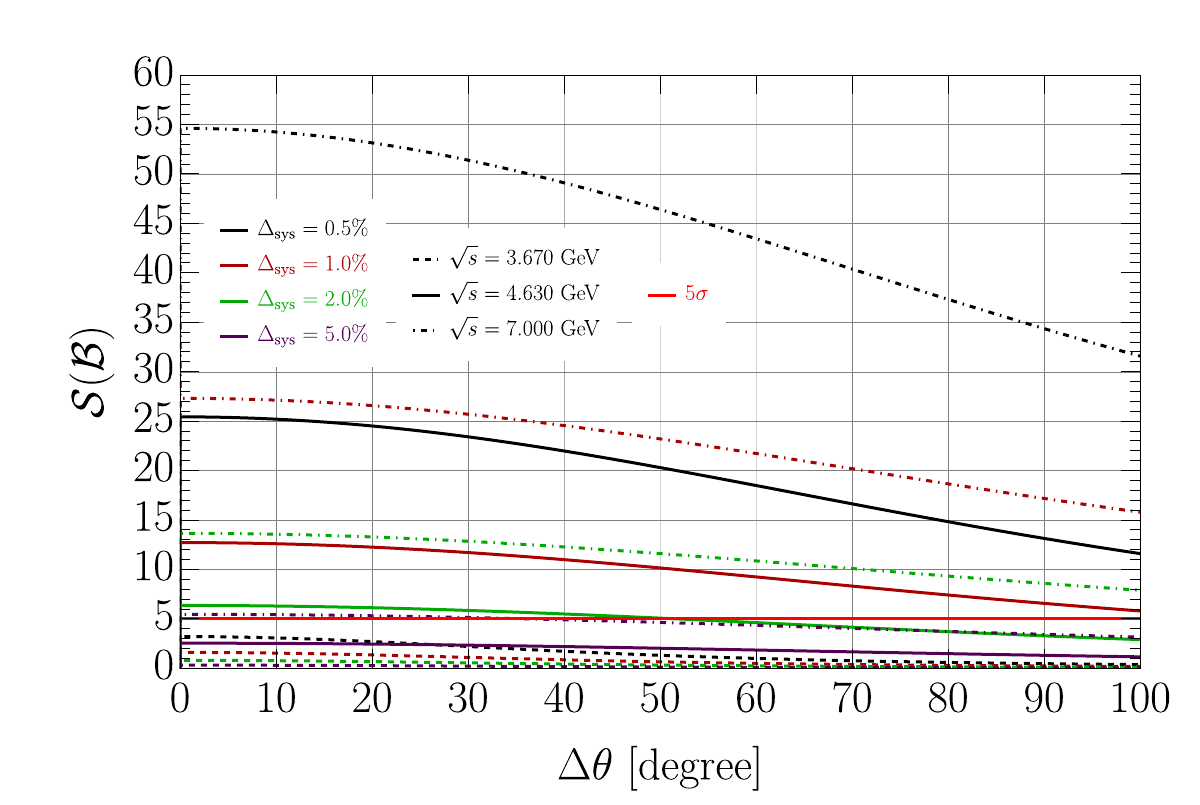}
    \caption{Projected statistical significances, $\mathcal{S}(\mathcal{C})$ (upper panel) and $\mathcal{S}(\mathcal{B})$ (lower panel), as functions of $\Delta\theta$, for different benchmark values of $\Delta_{\text{sys}}$ and COM energies. An integrated luminosity of $1~\mathrm{ab}^{-1}$ is assumed. The horizontal red line indicates the $5\sigma$ level.}
    \label{fig:SB_SC}
\end{figure}
Fig.~\ref{fig:SB_SC} shows the projected statistical significances as functions of $\Delta\theta$ for different choices of $\Delta_{\text{sys}}$ and $\sqrt{s}$. 
The significances increase as the angular window narrows and attain their largest values for $\Delta\theta$ slightly above $0^\circ$.\footnote{For $\Delta\theta=0$, no signal events remain.}
Such small angular windows are, however, below the expected angular resolution of the STCF main detector. 
The angular resolution of the STCF tracking system is expected to be a few mrad, corresponding to $\lesssim 1^\circ$~\cite{Achasov:2023gey}; considering the deteriorating effect of the missing neutrino in each $\tau$ decay, we take the conservative benchmark angular window of $10^\circ$ as the optimal case for the numerical evaluation below.
Furthermore, the projected statistical sensitivities increase with higher COM energies, reflecting the larger boost of the $\tau$-leptons and the enhanced spin correlations.

\begin{table}[t]
\centering
\begin{tabular}{|c | c | c | c | c |} 
 \hline
 $\Delta_{\text{sys}}$ & $\overline{\mathcal{C}}$ & $\Delta\mathcal{C}_{\text{tot}}$ & $\mathcal{S}(\mathcal{C})$ & $\Delta\theta$ \\ [0.5ex] 
 \hline
 0.5\% & 0.0320 & 0.0025 & $>5\sigma (8\%)$ & $\gtrsim 0^\circ(10^\circ)$\\
 1\%& 0.0320 & 0.0050 & $>5\sigma (16\%)$ & $\gtrsim 0^\circ(10^\circ)$\\
 2\% & 0.0320 & 0.0100 & $3.21\sigma$ & $\gtrsim 0^\circ(10^\circ)$\\
 5\% & 0.0320 & 0.0250 & $1.28\sigma$ & $\gtrsim 0^\circ(10^\circ)$\\ [1ex] 
 \hline
 \hline
  $\Delta_{\text{sys}}$ & $\overline{\mathcal{B}}-2$ & $\Delta\mathcal{B}$ & $\mathcal{S}(\mathcal{B})$ & $\Delta\theta$\\ [0.5ex]
  \hline
  0.5\% & 0.0010 & 0.0003 & $3.07\sigma$ & $\gtrsim 0^\circ(10^\circ)$ \\
  1\% &0.0010 & 0.0007 & $1.54\sigma$ & $\gtrsim 0^\circ(10^\circ)$ \\
  2\%& 0.0010 & 0.0013 & $0.77\sigma$ & $\gtrsim 0^\circ(10^\circ)$ \\
  5\% & 0.0010 & 0.0033 & $0.31\sigma$ & $\gtrsim 0^\circ(10^\circ)$ \\ [1ex]
  \hline
\end{tabular}
\caption{Projected statistical significances $\mathcal{S}(\mathcal{C})$ and $\mathcal{S}(\mathcal{B})$ for the $\sqrt{s}=3.670$~GeV dataset with an integrated luminosity of $1~\mathrm{ab}^{-1}$ at STCF, using the kinematic method. For cases where $\mathcal{S}>5\sigma$, the relative precision is indicated by $\mathcal{S}^{-1}$ in parentheses. 
While the analytically optimal values of $\Delta\theta$ are slightly above $0^\circ$, we take a realistic benchmark of $\Delta\theta=10^\circ$ for the numerical evaluation.}
\label{table:data_3p670}
\end{table}

\begin{table}[t]
\centering
\begin{tabular}{|c | c | c | c | c |} 
 \hline
 $\Delta_{\text{sys}}$ & $\overline{\mathcal{C}}$ & $\Delta\mathcal{C}_{\text{tot}}$ & $\mathcal{S}(\mathcal{C})$ & $\Delta\theta$ \\ [0.5ex] 
 \hline
 0.5\% & 0.2577 & 0.0027 & $>5\sigma (1\%)$ & $\gtrsim 0^\circ(10^\circ)$\\
 1\%& 0.2577 & 0.0053 & $>5\sigma (2\%)$ & $\gtrsim 0^\circ(10^\circ)$\\
 2\% & 0.2577 & 0.0106 & $>5\sigma (4\%)$ & $\gtrsim 0^\circ(10^\circ)$\\
 5\% & 0.2577 & 0.0266 & $>5\sigma (10\%)$ & $\gtrsim 0^\circ(10^\circ)$\\ [1ex] 
 \hline
 \hline
  $\Delta_{\text{sys}}$ & $\overline{\mathcal{B}}-2$ & $\Delta\mathcal{B}$ & $\mathcal{S}(\mathcal{B})$ & $\Delta\theta$\\ [0.5ex]
  \hline
  0.5\% & 0.0653 & 0.0026 & $>5\sigma (4\%)$ & $\gtrsim 0^\circ(10^\circ)$ \\
  1\% &0.0653 & 0.0052 & $>5\sigma (8\%)$ & $\gtrsim 0^\circ(10^\circ)$ \\
  2\%& 0.0653 & 0.0104 & $>5\sigma (16\%)$ & $\gtrsim 0^\circ(10^\circ)$ \\
  5\% & 0.0653 & 0.0259 & 2.52$\sigma$ & $\gtrsim 0^\circ(10^\circ)$ \\ [1ex]
  \hline
\end{tabular}
\caption{Projected statistical significances $\mathcal{S}(\mathcal{C})$ and $\mathcal{S}(\mathcal{B})$ for the $\sqrt{s}=4.630$~GeV dataset with an integrated luminosity of $1~\mathrm{ab}^{-1}$ at STCF.}
\label{table:data_4p630}
\end{table}

\begin{table}[t]
\centering
\begin{tabular}{|c | c | c | c | c |} 
 \hline
 $\Delta_{\text{sys}}$ & $\overline{\mathcal{C}}$ & $\Delta\mathcal{C}_{\text{tot}}$ & $\mathcal{S}(\mathcal{C})$ & $\Delta\theta$ \\ [0.5ex] 
 \hline
 0.5\% & 0.5877 & 0.0032 & $>5\sigma (1\%)$ & $\gtrsim 0^\circ(10^\circ)$\\
 1\%& 0.5877 & 0.0065 & $>5\sigma (1\%)$ & $\gtrsim 0^\circ(10^\circ)$\\
 2\% & 0.5877 & 0.0130 & $>5\sigma (2\%)$ & $\gtrsim 0^\circ(10^\circ)$\\
 5\% & 0.5877 & 0.0325 & $>5\sigma (6\%)$ & $\gtrsim 0^\circ(10^\circ)$\\ [1ex] 
 \hline
 \hline
  $\Delta_{\text{sys}}$ & $\overline{\mathcal{B}}-2$ & $\Delta\mathcal{B}$ & $\mathcal{S}(\mathcal{B})$ & $\Delta\theta$\\ [0.5ex]
  \hline
  0.5\% & 0.3199 & 0.0059 & $>5\sigma (2\%)$ & $\gtrsim 0^\circ(10^\circ)$ \\
  1\% &0.3199 & 0.0118 & $>5\sigma (4\%)$ & $\gtrsim 0^\circ(10^\circ)$ \\
  2\%& 0.3199 & 0.0236 & $>5\sigma (7\%)$ & $\gtrsim 0^\circ(10^\circ)$ \\
  5\% & 0.3199 & 0.0590 & $>5\sigma (18\%)$ & $\gtrsim 0^\circ(10^\circ)$ \\ [1ex]
  \hline
\end{tabular}\caption{Projected statistical significances $\mathcal{S}(\mathcal{C})$ and $\mathcal{S}(\mathcal{B})$ for the $\sqrt{s}=7.000$~GeV dataset with an integrated luminosity of $1~\mathrm{ab}^{-1}$ at STCF.}
\label{table:data_7p000}
\end{table}

Tables~\ref{table:data_3p670}--\ref{table:data_7p000} summarize the projected statistical significances $\mathcal{S}(\mathcal{C})$ and $\mathcal{S}(\mathcal{B})$ for the $3.670$, $4.630$, and $7.000$~GeV datasets with an integrated luminosity of $1~\mathrm{ab}^{-1}$ at STCF. 
The results indicate that higher COM energies lead to enhanced statistical sensitivities to both the concurrence and the Bell-type correlation observable, consistent with Eq.~\eqref{eqn:CandB} and Fig.~\ref{fig:B_C}. 

For $\sqrt{s}=3.670$~GeV, a $5\sigma$ sensitivity to the concurrence is only achievable if the relative systematic uncertainty is below about $1\%$, whereas the Bell-type spin-correlation observable does not reach $5\sigma$ even for $\Delta_{\text{sys}}=0.5\%$.
For $\sqrt{s}=4.630$~GeV, a $5\sigma$ sensitivity to the concurrence can be obtained for $\Delta_{\text{sys}}\lesssim 5\%$, corresponding to a percent-level relative precision.
The Bell-type correlation observable exhibits smaller significances, but values above $5\sigma$ are achieved provided $\Delta_{\text{sys}}\lesssim 2\%$ under the present assumptions.
For $\sqrt{s}=7.000$~GeV, statistical significances exceeding $5\sigma$ are projected for both $\mathcal{C}$ and $\mathcal{B}$ even for $\Delta_{\text{sys}}=5\%$.

We emphasize that detector-resolution effects are not included in the present analysis and could become relevant in regimes of limited statistics. 
In the numerical results shown above, an integrated luminosity of $1~\mathrm{ab}^{-1}$ is assumed. 
In this high-luminosity regime, the statistical uncertainties, which scale as $1/\sqrt{N}$ (see Eqs.~\eqref{eqn:DeltaCstat} and~\eqref{eqn:DeltaBstat}), are subdominant compared to the assumed systematic uncertainties. 
As a result, the projected significances are largely insensitive to further increases in the collected data volume within this setup.

In contrast, for substantially smaller integrated luminosities, the statistical uncertainties are expected to dominate over the systematic ones. Within the present kinematic approach, however, the statistical variance is evaluated through phase-space integration over the production angle under the assumption of ideal event reconstruction.
In the absence of detector smearing effects, this procedure can yield very small values of the variance $\mathrm{Var}$ in narrowly selected angular regions, and consequently statistical significances that remain nearly unchanged even for $N_{\mathrm{tot}} = 10$--$100$. This feature should be regarded as an artifact of the idealized setup considered here, rather than as an indication that such limited event samples would suffice to reach the $5\sigma$ level in a realistic experimental analysis.

\section{Conclusions}\label{sec:conclusions}

In this paper, we have investigated the projected sensitivity to entanglement measures and Bell-type spin-correlation observables in the electroweak process $e^- e^+\to \tau^-\tau^+$ at the Super Tau-Charm Facility.

We applied a kinematic method that determines the relevant spin-correlation coefficients from the $\tau$-lepton velocity $\beta$ and scattering angle $\theta$ in the COM frame. 
The concurrence $\mathcal{C}$ and the Bell-type correlation variable $\mathcal{B}$ were defined in terms of the Fano coefficients of the density matrix within the Standard-Model framework. 
These coefficients were computed as functions of $\beta$ and $\theta$. 
Focusing on the decay channel $\tau^\pm \to \pi^\pm \nu_\tau$, we imposed angular cuts in the window $[\pi/2-\Delta\theta/2, \pi/2+\Delta\theta/2]$ and estimated the corresponding signal yields. 
Including statistical uncertainties and assuming representative levels of systematic uncertainty, we evaluated the projected statistical significances $\mathcal{S}(\mathcal{C})$ and $\mathcal{S}(\mathcal{B})$ for $\sqrt{s}=3.670$, $4.630$, and $7.000$~GeV.

Our numerical results indicate that percent-level precision on $\mathcal{C}$ is attainable for relative systematic uncertainties $\Delta_{\text{sys}}\lesssim 5\%$ at $\sqrt{s}=4.630$ and $7.000$~GeV under the present assumptions. 
For the Bell-type correlation observable, statistical significances above the $5\sigma$ level are projected for $\Delta_{\text{sys}}\lesssim 2\%$ ($5\%$) at $\sqrt{s}=4.630$ ($7.000$)~GeV. 
Compared with BESIII, the enhanced sensitivity at STCF primarily arises from its higher COM energies and, secondarily, from its substantially larger integrated luminosity. 
Finally, we note that a realistic detector-level simulation will be necessary to reliably determine the minimal integrated luminosity required to reach the quoted significance levels in an experimental analysis.

\section*{Acknowledgments}

We are grateful to Herbi Dreiner for carefully reading the manuscript and providing insightful comments.
We thank Mingyi Liu and Youle Su for useful discussions on the estimated polar-angle resolution of the reconstructed $\tau$-leptons at STCF and the employed computation procedures of the kinematic method, respectively.
This work was supported by the National Natural Science Foundation of China under grant Nos.~12475106 and~12505120, the Fundamental Research Funds for the Central Universities under Grant No.~JZ2025HGTG0252, and the CAS Youth Team Program under Contract No.~YSBR-101.



\bibliography{refs}

\begin{thebibliography}{128}%
\makeatletter
\providecommand \@ifxundefined [1]{%
 \@ifx{#1\undefined}
}%
\providecommand \@ifnum [1]{%
 \ifnum #1\expandafter \@firstoftwo
 \else \expandafter \@secondoftwo
 \fi
}%
\providecommand \@ifx [1]{%
 \ifx #1\expandafter \@firstoftwo
 \else \expandafter \@secondoftwo
 \fi
}%
\providecommand \natexlab [1]{#1}%
\providecommand \enquote  [1]{``#1''}%
\providecommand \bibnamefont  [1]{#1}%
\providecommand \bibfnamefont [1]{#1}%
\providecommand \citenamefont [1]{#1}%
\providecommand \href@noop [0]{\@secondoftwo}%
\providecommand \href [0]{\begingroup \@sanitize@url \@href}%
\providecommand \@href[1]{\@@startlink{#1}\@@href}%
\providecommand \@@href[1]{\endgroup#1\@@endlink}%
\providecommand \@sanitize@url [0]{\catcode `\\12\catcode `\$12\catcode
  `\&12\catcode `\#12\catcode `\^12\catcode `\_12\catcode `\%12\relax}%
\providecommand \@@startlink[1]{}%
\providecommand \@@endlink[0]{}%
\providecommand \url  [0]{\begingroup\@sanitize@url \@url }%
\providecommand \@url [1]{\endgroup\@href {#1}{\urlprefix }}%
\providecommand \urlprefix  [0]{URL }%
\providecommand \Eprint [0]{\href }%
\providecommand \doibase [0]{http://dx.doi.org/}%
\providecommand \selectlanguage [0]{\@gobble}%
\providecommand \bibinfo  [0]{\@secondoftwo}%
\providecommand \bibfield  [0]{\@secondoftwo}%
\providecommand \translation [1]{[#1]}%
\providecommand \BibitemOpen [0]{}%
\providecommand \bibitemStop [0]{}%
\providecommand \bibitemNoStop [0]{.\EOS\space}%
\providecommand \EOS [0]{\spacefactor3000\relax}%
\providecommand \BibitemShut  [1]{\csname bibitem#1\endcsname}%
\let\auto@bib@innerbib\@empty
\bibitem [{\citenamefont {Brunner}\ \emph {et~al.}(2014)\citenamefont
  {Brunner}, \citenamefont {Cavalcanti}, \citenamefont {Pironio}, \citenamefont
  {Scarani},\ and\ \citenamefont {Wehner}}]{Brunner:2013est}%
  \BibitemOpen
  \bibfield  {author} {\bibinfo {author} {\bibfnamefont {N.}~\bibnamefont
  {Brunner}}, \bibinfo {author} {\bibfnamefont {D.}~\bibnamefont {Cavalcanti}},
  \bibinfo {author} {\bibfnamefont {S.}~\bibnamefont {Pironio}}, \bibinfo
  {author} {\bibfnamefont {V.}~\bibnamefont {Scarani}}, \ and\ \bibinfo
  {author} {\bibfnamefont {S.}~\bibnamefont {Wehner}},\ }\href {\doibase
  10.1103/RevModPhys.86.419} {\bibfield  {journal} {\bibinfo  {journal} {Rev.
  Mod. Phys.}\ }\textbf {\bibinfo {volume} {86}},\ \bibinfo {pages} {419}
  (\bibinfo {year} {2014})},\ \Eprint {http://arxiv.org/abs/1303.2849}
  {arXiv:1303.2849 [quant-ph]} \BibitemShut {NoStop}%
\bibitem [{\citenamefont {Horodecki}\ \emph {et~al.}(2009)\citenamefont
  {Horodecki}, \citenamefont {Horodecki}, \citenamefont {Horodecki},\ and\
  \citenamefont {Horodecki}}]{Horodecki:2009zz}%
  \BibitemOpen
  \bibfield  {author} {\bibinfo {author} {\bibfnamefont {R.}~\bibnamefont
  {Horodecki}}, \bibinfo {author} {\bibfnamefont {P.}~\bibnamefont
  {Horodecki}}, \bibinfo {author} {\bibfnamefont {M.}~\bibnamefont
  {Horodecki}}, \ and\ \bibinfo {author} {\bibfnamefont {K.}~\bibnamefont
  {Horodecki}},\ }\href {\doibase 10.1103/RevModPhys.81.865} {\bibfield
  {journal} {\bibinfo  {journal} {Rev. Mod. Phys.}\ }\textbf {\bibinfo {volume}
  {81}},\ \bibinfo {pages} {865} (\bibinfo {year} {2009})},\ \Eprint
  {http://arxiv.org/abs/quant-ph/0702225} {arXiv:quant-ph/0702225} \BibitemShut
  {NoStop}%
\bibitem [{\citenamefont {Bell}(1964)}]{Bell:1964kc}%
  \BibitemOpen
  \bibfield  {author} {\bibinfo {author} {\bibfnamefont {J.~S.}\ \bibnamefont
  {Bell}},\ }\href {\doibase 10.1103/PhysicsPhysiqueFizika.1.195} {\bibfield
  {journal} {\bibinfo  {journal} {Physics Physique Fizika}\ }\textbf {\bibinfo
  {volume} {1}},\ \bibinfo {pages} {195} (\bibinfo {year} {1964})}\BibitemShut
  {NoStop}%
\bibitem [{\citenamefont {Clauser}\ and\ \citenamefont
  {Horne}(1974)}]{Clauser:1974tg}%
  \BibitemOpen
  \bibfield  {author} {\bibinfo {author} {\bibfnamefont {J.~F.}\ \bibnamefont
  {Clauser}}\ and\ \bibinfo {author} {\bibfnamefont {M.~A.}\ \bibnamefont
  {Horne}},\ }\href {\doibase 10.1103/PhysRevD.10.526} {\bibfield  {journal}
  {\bibinfo  {journal} {Phys. Rev. D}\ }\textbf {\bibinfo {volume} {10}},\
  \bibinfo {pages} {526} (\bibinfo {year} {1974})}\BibitemShut {NoStop}%
\bibitem [{\citenamefont {Aspect}\ \emph {et~al.}(1981)\citenamefont {Aspect},
  \citenamefont {Grangier},\ and\ \citenamefont {Roger}}]{Aspect:1981zz}%
  \BibitemOpen
  \bibfield  {author} {\bibinfo {author} {\bibfnamefont {A.}~\bibnamefont
  {Aspect}}, \bibinfo {author} {\bibfnamefont {P.}~\bibnamefont {Grangier}}, \
  and\ \bibinfo {author} {\bibfnamefont {G.}~\bibnamefont {Roger}},\ }\href
  {\doibase 10.1103/PhysRevLett.47.460} {\bibfield  {journal} {\bibinfo
  {journal} {Phys. Rev. Lett.}\ }\textbf {\bibinfo {volume} {47}},\ \bibinfo
  {pages} {460} (\bibinfo {year} {1981})}\BibitemShut {NoStop}%
\bibitem [{\citenamefont {Bohm}\ and\ \citenamefont
  {Aharonov}(1957)}]{Bohm:1957zz}%
  \BibitemOpen
  \bibfield  {author} {\bibinfo {author} {\bibfnamefont {D.}~\bibnamefont
  {Bohm}}\ and\ \bibinfo {author} {\bibfnamefont {Y.}~\bibnamefont
  {Aharonov}},\ }\href {\doibase 10.1103/PhysRev.108.1070} {\bibfield
  {journal} {\bibinfo  {journal} {Phys. Rev.}\ }\textbf {\bibinfo {volume}
  {108}},\ \bibinfo {pages} {1070} (\bibinfo {year} {1957})}\BibitemShut
  {NoStop}%
\bibitem [{\citenamefont {Einstein}\ \emph {et~al.}(1935)\citenamefont
  {Einstein}, \citenamefont {Podolsky},\ and\ \citenamefont
  {Rosen}}]{Einstein:1935rr}%
  \BibitemOpen
  \bibfield  {author} {\bibinfo {author} {\bibfnamefont {A.}~\bibnamefont
  {Einstein}}, \bibinfo {author} {\bibfnamefont {B.}~\bibnamefont {Podolsky}},
  \ and\ \bibinfo {author} {\bibfnamefont {N.}~\bibnamefont {Rosen}},\ }\href
  {\doibase 10.1103/PhysRev.47.777} {\bibfield  {journal} {\bibinfo  {journal}
  {Phys. Rev.}\ }\textbf {\bibinfo {volume} {47}},\ \bibinfo {pages} {777}
  (\bibinfo {year} {1935})}\BibitemShut {NoStop}%
\bibitem [{\citenamefont {Freedman}\ and\ \citenamefont
  {Clauser}(1972)}]{Freedman:1972zza}%
  \BibitemOpen
  \bibfield  {author} {\bibinfo {author} {\bibfnamefont {S.~J.}\ \bibnamefont
  {Freedman}}\ and\ \bibinfo {author} {\bibfnamefont {J.~F.}\ \bibnamefont
  {Clauser}},\ }\href {\doibase 10.1103/PhysRevLett.28.938} {\bibfield
  {journal} {\bibinfo  {journal} {Phys. Rev. Lett.}\ }\textbf {\bibinfo
  {volume} {28}},\ \bibinfo {pages} {938} (\bibinfo {year} {1972})}\BibitemShut
  {NoStop}%
\bibitem [{\citenamefont {Aspect}\ \emph
  {et~al.}(1982{\natexlab{a}})\citenamefont {Aspect}, \citenamefont
  {Grangier},\ and\ \citenamefont {Roger}}]{Aspect:1981nv}%
  \BibitemOpen
  \bibfield  {author} {\bibinfo {author} {\bibfnamefont {A.}~\bibnamefont
  {Aspect}}, \bibinfo {author} {\bibfnamefont {P.}~\bibnamefont {Grangier}}, \
  and\ \bibinfo {author} {\bibfnamefont {G.}~\bibnamefont {Roger}},\ }\href
  {\doibase 10.1103/PhysRevLett.49.91} {\bibfield  {journal} {\bibinfo
  {journal} {Phys. Rev. Lett.}\ }\textbf {\bibinfo {volume} {49}},\ \bibinfo
  {pages} {91} (\bibinfo {year} {1982}{\natexlab{a}})}\BibitemShut {NoStop}%
\bibitem [{\citenamefont {Aspect}\ \emph
  {et~al.}(1982{\natexlab{b}})\citenamefont {Aspect}, \citenamefont
  {Dalibard},\ and\ \citenamefont {Roger}}]{Aspect:1982fx}%
  \BibitemOpen
  \bibfield  {author} {\bibinfo {author} {\bibfnamefont {A.}~\bibnamefont
  {Aspect}}, \bibinfo {author} {\bibfnamefont {J.}~\bibnamefont {Dalibard}}, \
  and\ \bibinfo {author} {\bibfnamefont {G.}~\bibnamefont {Roger}},\ }\href
  {\doibase 10.1103/PhysRevLett.49.1804} {\bibfield  {journal} {\bibinfo
  {journal} {Phys. Rev. Lett.}\ }\textbf {\bibinfo {volume} {49}},\ \bibinfo
  {pages} {1804} (\bibinfo {year} {1982}{\natexlab{b}})}\BibitemShut {NoStop}%
\bibitem [{\citenamefont {Weihs}\ \emph {et~al.}(1998)\citenamefont {Weihs},
  \citenamefont {Jennewein}, \citenamefont {Simon}, \citenamefont
  {Weinfurter},\ and\ \citenamefont {Zeilinger}}]{Weihs:1998gy}%
  \BibitemOpen
  \bibfield  {author} {\bibinfo {author} {\bibfnamefont {G.}~\bibnamefont
  {Weihs}}, \bibinfo {author} {\bibfnamefont {T.}~\bibnamefont {Jennewein}},
  \bibinfo {author} {\bibfnamefont {C.}~\bibnamefont {Simon}}, \bibinfo
  {author} {\bibfnamefont {H.}~\bibnamefont {Weinfurter}}, \ and\ \bibinfo
  {author} {\bibfnamefont {A.}~\bibnamefont {Zeilinger}},\ }\href {\doibase
  10.1103/PhysRevLett.81.5039} {\bibfield  {journal} {\bibinfo  {journal}
  {Phys. Rev. Lett.}\ }\textbf {\bibinfo {volume} {81}},\ \bibinfo {pages}
  {5039} (\bibinfo {year} {1998})},\ \Eprint
  {http://arxiv.org/abs/quant-ph/9810080} {arXiv:quant-ph/9810080} \BibitemShut
  {NoStop}%
\bibitem [{\citenamefont {Hagley}\ \emph {et~al.}(1997)\citenamefont {Hagley},
  \citenamefont {Ma{\^\i}tre}, \citenamefont {Nogues}, \citenamefont
  {Wunderlich}, \citenamefont {Brune}, \citenamefont {Raimond},\ and\
  \citenamefont {Haroche}}]{Hagley:1997bob}%
  \BibitemOpen
  \bibfield  {author} {\bibinfo {author} {\bibfnamefont {E.}~\bibnamefont
  {Hagley}}, \bibinfo {author} {\bibfnamefont {X.}~\bibnamefont {Ma{\^\i}tre}},
  \bibinfo {author} {\bibfnamefont {G.}~\bibnamefont {Nogues}}, \bibinfo
  {author} {\bibfnamefont {C.}~\bibnamefont {Wunderlich}}, \bibinfo {author}
  {\bibfnamefont {M.}~\bibnamefont {Brune}}, \bibinfo {author} {\bibfnamefont
  {J.~M.}\ \bibnamefont {Raimond}}, \ and\ \bibinfo {author} {\bibfnamefont
  {S.}~\bibnamefont {Haroche}},\ }\href {\doibase 10.1103/PhysRevLett.79.1}
  {\bibfield  {journal} {\bibinfo  {journal} {Phys. Rev. Lett.}\ }\textbf
  {\bibinfo {volume} {79}},\ \bibinfo {pages} {1} (\bibinfo {year}
  {1997})}\BibitemShut {NoStop}%
\bibitem [{\citenamefont {Yin}\ \emph {et~al.}(2017)\citenamefont {Yin} \emph
  {et~al.}}]{Yin:2017ips}%
  \BibitemOpen
  \bibfield  {author} {\bibinfo {author} {\bibfnamefont {J.}~\bibnamefont
  {Yin}} \emph {et~al.},\ }\href {\doibase 10.1126/science.aan3211} {\bibfield
  {journal} {\bibinfo  {journal} {Science}\ }\textbf {\bibinfo {volume}
  {356}},\ \bibinfo {pages} {aan3211} (\bibinfo {year} {2017})}\BibitemShut
  {NoStop}%
\bibitem [{\citenamefont {Abell{\'a}n}\ \emph {et~al.}(2018)\citenamefont
  {Abell{\'a}n} \emph {et~al.}}]{BIGBellTest:2018ebd}%
  \BibitemOpen
  \bibfield  {author} {\bibinfo {author} {\bibfnamefont {C.}~\bibnamefont
  {Abell{\'a}n}} \emph {et~al.} (\bibinfo {collaboration} {BIG Bell Test}),\
  }\href {\doibase 10.1038/s41586-018-0085-3} {\bibfield  {journal} {\bibinfo
  {journal} {Nature}\ }\textbf {\bibinfo {volume} {557}},\ \bibinfo {pages}
  {212} (\bibinfo {year} {2018})},\ \Eprint {http://arxiv.org/abs/1805.04431}
  {arXiv:1805.04431 [quant-ph]} \BibitemShut {NoStop}%
\bibitem [{\citenamefont {Clauser}\ and\ \citenamefont
  {Shimony}(1978)}]{Clauser:1978ng}%
  \BibitemOpen
  \bibfield  {author} {\bibinfo {author} {\bibfnamefont {J.~F.}\ \bibnamefont
  {Clauser}}\ and\ \bibinfo {author} {\bibfnamefont {A.}~\bibnamefont
  {Shimony}},\ }\href {\doibase 10.1088/0034-4885/41/12/002} {\bibfield
  {journal} {\bibinfo  {journal} {Rept. Prog. Phys.}\ }\textbf {\bibinfo
  {volume} {41}},\ \bibinfo {pages} {1881} (\bibinfo {year}
  {1978})}\BibitemShut {NoStop}%
\bibitem [{\citenamefont {Genovese}(2005)}]{Genovese:2005nw}%
  \BibitemOpen
  \bibfield  {author} {\bibinfo {author} {\bibfnamefont {M.}~\bibnamefont
  {Genovese}},\ }\href {\doibase 10.1016/j.physrep.2005.03.003} {\bibfield
  {journal} {\bibinfo  {journal} {Phys. Rept.}\ }\textbf {\bibinfo {volume}
  {413}},\ \bibinfo {pages} {319} (\bibinfo {year} {2005})},\ \Eprint
  {http://arxiv.org/abs/quant-ph/0701071} {arXiv:quant-ph/0701071} \BibitemShut
  {NoStop}%
\bibitem [{\citenamefont {Tornqvist}(1981)}]{Tornqvist:1980af}%
  \BibitemOpen
  \bibfield  {author} {\bibinfo {author} {\bibfnamefont {N.~A.}\ \bibnamefont
  {Tornqvist}},\ }\href {\doibase 10.1007/BF00715204} {\bibfield  {journal}
  {\bibinfo  {journal} {Found. Phys.}\ }\textbf {\bibinfo {volume} {11}},\
  \bibinfo {pages} {171} (\bibinfo {year} {1981})}\BibitemShut {NoStop}%
\bibitem [{\citenamefont {Abel}\ \emph {et~al.}(1992)\citenamefont {Abel},
  \citenamefont {Dittmar},\ and\ \citenamefont {Dreiner}}]{Abel:1992kz}%
  \BibitemOpen
  \bibfield  {author} {\bibinfo {author} {\bibfnamefont {S.~A.}\ \bibnamefont
  {Abel}}, \bibinfo {author} {\bibfnamefont {M.}~\bibnamefont {Dittmar}}, \
  and\ \bibinfo {author} {\bibfnamefont {H.~K.}\ \bibnamefont {Dreiner}},\
  }\href {\doibase 10.1016/0370-2693(92)90071-B} {\bibfield  {journal}
  {\bibinfo  {journal} {Phys. Lett. B}\ }\textbf {\bibinfo {volume} {280}},\
  \bibinfo {pages} {304} (\bibinfo {year} {1992})}\BibitemShut {NoStop}%
\bibitem [{\citenamefont {Benatti}\ and\ \citenamefont
  {Floreanini}(1998)}]{Benatti:1997fr}%
  \BibitemOpen
  \bibfield  {author} {\bibinfo {author} {\bibfnamefont {F.}~\bibnamefont
  {Benatti}}\ and\ \bibinfo {author} {\bibfnamefont {R.}~\bibnamefont
  {Floreanini}},\ }\href {\doibase 10.1103/PhysRevD.57.R1332} {\bibfield
  {journal} {\bibinfo  {journal} {Phys. Rev. D}\ }\textbf {\bibinfo {volume}
  {57}},\ \bibinfo {pages} {R1332} (\bibinfo {year} {1998})},\ \Eprint
  {http://arxiv.org/abs/hep-ph/9712274} {arXiv:hep-ph/9712274} \BibitemShut
  {NoStop}%
\bibitem [{\citenamefont {Benatti}\ and\ \citenamefont
  {Floreanini}(1999)}]{Benatti:1999jt}%
  \BibitemOpen
  \bibfield  {author} {\bibinfo {author} {\bibfnamefont {F.}~\bibnamefont
  {Benatti}}\ and\ \bibinfo {author} {\bibfnamefont {R.}~\bibnamefont
  {Floreanini}},\ }\href {\doibase 10.1142/S0217732399001619} {\bibfield
  {journal} {\bibinfo  {journal} {Mod. Phys. Lett. A}\ }\textbf {\bibinfo
  {volume} {14}},\ \bibinfo {pages} {1519} (\bibinfo {year} {1999})},\ \Eprint
  {http://arxiv.org/abs/hep-ph/9906272} {arXiv:hep-ph/9906272} \BibitemShut
  {NoStop}%
\bibitem [{\citenamefont {Benatti}\ and\ \citenamefont
  {Floreanini}(2000)}]{Benatti:1999du}%
  \BibitemOpen
  \bibfield  {author} {\bibinfo {author} {\bibfnamefont {F.}~\bibnamefont
  {Benatti}}\ and\ \bibinfo {author} {\bibfnamefont {R.}~\bibnamefont
  {Floreanini}},\ }\href {\doibase 10.1007/s100520050692} {\bibfield  {journal}
  {\bibinfo  {journal} {Eur. Phys. J. C}\ }\textbf {\bibinfo {volume} {13}},\
  \bibinfo {pages} {267} (\bibinfo {year} {2000})},\ \Eprint
  {http://arxiv.org/abs/hep-ph/9912348} {arXiv:hep-ph/9912348} \BibitemShut
  {NoStop}%
\bibitem [{\citenamefont {Bertlmann}\ \emph {et~al.}(2001)\citenamefont
  {Bertlmann}, \citenamefont {Grimus},\ and\ \citenamefont
  {Hiesmayr}}]{Bertlmann:2001ea}%
  \BibitemOpen
  \bibfield  {author} {\bibinfo {author} {\bibfnamefont {R.~A.}\ \bibnamefont
  {Bertlmann}}, \bibinfo {author} {\bibfnamefont {W.}~\bibnamefont {Grimus}}, \
  and\ \bibinfo {author} {\bibfnamefont {B.~C.}\ \bibnamefont {Hiesmayr}},\
  }\href {\doibase 10.1016/S0375-9601(01)00577-1} {\bibfield  {journal}
  {\bibinfo  {journal} {Phys. Lett. A}\ }\textbf {\bibinfo {volume} {289}},\
  \bibinfo {pages} {21} (\bibinfo {year} {2001})},\ \Eprint
  {http://arxiv.org/abs/quant-ph/0107022} {arXiv:quant-ph/0107022} \BibitemShut
  {NoStop}%
\bibitem [{\citenamefont {Banerjee}\ \emph {et~al.}(2016)\citenamefont
  {Banerjee}, \citenamefont {Alok},\ and\ \citenamefont
  {MacKenzie}}]{Banerjee:2014vga}%
  \BibitemOpen
  \bibfield  {author} {\bibinfo {author} {\bibfnamefont {S.}~\bibnamefont
  {Banerjee}}, \bibinfo {author} {\bibfnamefont {A.~K.}\ \bibnamefont {Alok}},
  \ and\ \bibinfo {author} {\bibfnamefont {R.}~\bibnamefont {MacKenzie}},\
  }\href {\doibase 10.1140/epjp/i2016-16129-0} {\bibfield  {journal} {\bibinfo
  {journal} {Eur. Phys. J. Plus}\ }\textbf {\bibinfo {volume} {131}},\ \bibinfo
  {pages} {129} (\bibinfo {year} {2016})},\ \Eprint
  {http://arxiv.org/abs/1409.1034} {arXiv:1409.1034 [hep-ph]} \BibitemShut
  {NoStop}%
\bibitem [{\citenamefont {Acin}\ \emph {et~al.}(2001)\citenamefont {Acin},
  \citenamefont {Latorre},\ and\ \citenamefont {Pascual}}]{Acin:2000cs}%
  \BibitemOpen
  \bibfield  {author} {\bibinfo {author} {\bibfnamefont {A.}~\bibnamefont
  {Acin}}, \bibinfo {author} {\bibfnamefont {J.~I.}\ \bibnamefont {Latorre}}, \
  and\ \bibinfo {author} {\bibfnamefont {P.}~\bibnamefont {Pascual}},\ }\href
  {\doibase 10.1103/PhysRevA.63.042107} {\bibfield  {journal} {\bibinfo
  {journal} {Phys. Rev. A}\ }\textbf {\bibinfo {volume} {63}},\ \bibinfo
  {pages} {042107} (\bibinfo {year} {2001})},\ \Eprint
  {http://arxiv.org/abs/quant-ph/0007080} {arXiv:quant-ph/0007080} \BibitemShut
  {NoStop}%
\bibitem [{\citenamefont {Li}\ and\ \citenamefont {Qiao}(2009)}]{Li:2008dk}%
  \BibitemOpen
  \bibfield  {author} {\bibinfo {author} {\bibfnamefont {J.}~\bibnamefont
  {Li}}\ and\ \bibinfo {author} {\bibfnamefont {C.-F.}\ \bibnamefont {Qiao}},\
  }\href {\doibase 10.1016/j.physleta.2009.09.057} {\bibfield  {journal}
  {\bibinfo  {journal} {Phys. Lett. A}\ }\textbf {\bibinfo {volume} {373}},\
  \bibinfo {pages} {4311} (\bibinfo {year} {2009})},\ \Eprint
  {http://arxiv.org/abs/0812.0869} {arXiv:0812.0869 [quant-ph]} \BibitemShut
  {NoStop}%
\bibitem [{\citenamefont {Baranov}(2008)}]{Baranov:2008zzb}%
  \BibitemOpen
  \bibfield  {author} {\bibinfo {author} {\bibfnamefont {S.~P.}\ \bibnamefont
  {Baranov}},\ }\href {\doibase 10.1088/0954-3899/35/7/075002} {\bibfield
  {journal} {\bibinfo  {journal} {J. Phys. G}\ }\textbf {\bibinfo {volume}
  {35}},\ \bibinfo {pages} {075002} (\bibinfo {year} {2008})}\BibitemShut
  {NoStop}%
\bibitem [{\citenamefont {Chen}\ \emph {et~al.}(2013)\citenamefont {Chen},
  \citenamefont {Nakaguchi},\ and\ \citenamefont {Komamiya}}]{Chen:2013epa}%
  \BibitemOpen
  \bibfield  {author} {\bibinfo {author} {\bibfnamefont {S.}~\bibnamefont
  {Chen}}, \bibinfo {author} {\bibfnamefont {Y.}~\bibnamefont {Nakaguchi}}, \
  and\ \bibinfo {author} {\bibfnamefont {S.}~\bibnamefont {Komamiya}},\ }\href
  {\doibase 10.1093/ptep/ptt032} {\bibfield  {journal} {\bibinfo  {journal}
  {PTEP}\ }\textbf {\bibinfo {volume} {2013}},\ \bibinfo {pages} {063A01}
  (\bibinfo {year} {2013})},\ \Eprint {http://arxiv.org/abs/1302.6438}
  {arXiv:1302.6438 [hep-ph]} \BibitemShut {NoStop}%
\bibitem [{\citenamefont {Qian}\ \emph {et~al.}(2020)\citenamefont {Qian},
  \citenamefont {Li}, \citenamefont {Khan},\ and\ \citenamefont
  {Qiao}}]{Qian:2020ini}%
  \BibitemOpen
  \bibfield  {author} {\bibinfo {author} {\bibfnamefont {C.}~\bibnamefont
  {Qian}}, \bibinfo {author} {\bibfnamefont {J.-L.}\ \bibnamefont {Li}},
  \bibinfo {author} {\bibfnamefont {A.~S.}\ \bibnamefont {Khan}}, \ and\
  \bibinfo {author} {\bibfnamefont {C.-F.}\ \bibnamefont {Qiao}},\ }\href
  {\doibase 10.1103/PhysRevD.101.116004} {\bibfield  {journal} {\bibinfo
  {journal} {Phys. Rev. D}\ }\textbf {\bibinfo {volume} {101}},\ \bibinfo
  {pages} {116004} (\bibinfo {year} {2020})},\ \Eprint
  {http://arxiv.org/abs/2002.04283} {arXiv:2002.04283 [quant-ph]} \BibitemShut
  {NoStop}%
\bibitem [{\citenamefont {Banerjee}\ \emph {et~al.}(2015)\citenamefont
  {Banerjee}, \citenamefont {Alok}, \citenamefont {Srikanth},\ and\
  \citenamefont {Hiesmayr}}]{Banerjee:2015mha}%
  \BibitemOpen
  \bibfield  {author} {\bibinfo {author} {\bibfnamefont {S.}~\bibnamefont
  {Banerjee}}, \bibinfo {author} {\bibfnamefont {A.~K.}\ \bibnamefont {Alok}},
  \bibinfo {author} {\bibfnamefont {R.}~\bibnamefont {Srikanth}}, \ and\
  \bibinfo {author} {\bibfnamefont {B.~C.}\ \bibnamefont {Hiesmayr}},\ }\href
  {\doibase 10.1140/epjc/s10052-015-3717-x} {\bibfield  {journal} {\bibinfo
  {journal} {Eur. Phys. J. C}\ }\textbf {\bibinfo {volume} {75}},\ \bibinfo
  {pages} {487} (\bibinfo {year} {2015})},\ \Eprint
  {http://arxiv.org/abs/1508.03480} {arXiv:1508.03480 [hep-ph]} \BibitemShut
  {NoStop}%
\bibitem [{\citenamefont {Yongram}\ and\ \citenamefont
  {Manoukian}(2013)}]{Yongram:2013soa}%
  \BibitemOpen
  \bibfield  {author} {\bibinfo {author} {\bibfnamefont {N.}~\bibnamefont
  {Yongram}}\ and\ \bibinfo {author} {\bibfnamefont {E.~B.}\ \bibnamefont
  {Manoukian}},\ }\href {\doibase 10.1002/prop.201200137} {\bibfield  {journal}
  {\bibinfo  {journal} {Fortsch. Phys.}\ }\textbf {\bibinfo {volume} {61}},\
  \bibinfo {pages} {668} (\bibinfo {year} {2013})},\ \Eprint
  {http://arxiv.org/abs/1309.2059} {arXiv:1309.2059 [hep-th]} \BibitemShut
  {NoStop}%
\bibitem [{\citenamefont {Cervera-Lierta}\ \emph {et~al.}(2017)\citenamefont
  {Cervera-Lierta}, \citenamefont {Latorre}, \citenamefont {Rojo},\ and\
  \citenamefont {Rottoli}}]{Cervera-Lierta:2017tdt}%
  \BibitemOpen
  \bibfield  {author} {\bibinfo {author} {\bibfnamefont {A.}~\bibnamefont
  {Cervera-Lierta}}, \bibinfo {author} {\bibfnamefont {J.~I.}\ \bibnamefont
  {Latorre}}, \bibinfo {author} {\bibfnamefont {J.}~\bibnamefont {Rojo}}, \
  and\ \bibinfo {author} {\bibfnamefont {L.}~\bibnamefont {Rottoli}},\ }\href
  {\doibase 10.21468/SciPostPhys.3.5.036} {\bibfield  {journal} {\bibinfo
  {journal} {SciPost Phys.}\ }\textbf {\bibinfo {volume} {3}},\ \bibinfo
  {pages} {036} (\bibinfo {year} {2017})},\ \Eprint
  {http://arxiv.org/abs/1703.02989} {arXiv:1703.02989 [hep-th]} \BibitemShut
  {NoStop}%
\bibitem [{\citenamefont {Aspect}(2002)}]{Aspect2002}%
  \BibitemOpen
  \bibfield  {author} {\bibinfo {author} {\bibfnamefont {A.}~\bibnamefont
  {Aspect}},\ }\enquote {\bibinfo {title} {Bell's theorem: The naive view of an
  experimentalist},}\ in\ \href {\doibase 10.1007/978-3-662-05032-3_9} {\emph
  {\bibinfo {booktitle} {Quantum [Un]speakables: From Bell to Quantum
  Information}}}\ (\bibinfo  {publisher} {Springer Berlin Heidelberg},\
  \bibinfo {address} {Berlin, Heidelberg},\ \bibinfo {year} {2002})\ pp.\
  \bibinfo {pages} {119--153}\BibitemShut {NoStop}%
\bibitem [{\citenamefont {Fabbrichesi}\ \emph
  {et~al.}(2024{\natexlab{a}})\citenamefont {Fabbrichesi}, \citenamefont
  {Floreanini}, \citenamefont {Gabrielli},\ and\ \citenamefont
  {Marzola}}]{Fabbrichesi:2023idl}%
  \BibitemOpen
  \bibfield  {author} {\bibinfo {author} {\bibfnamefont {M.}~\bibnamefont
  {Fabbrichesi}}, \bibinfo {author} {\bibfnamefont {R.}~\bibnamefont
  {Floreanini}}, \bibinfo {author} {\bibfnamefont {E.}~\bibnamefont
  {Gabrielli}}, \ and\ \bibinfo {author} {\bibfnamefont {L.}~\bibnamefont
  {Marzola}},\ }\href {\doibase 10.1103/PhysRevD.109.L031104} {\bibfield
  {journal} {\bibinfo  {journal} {Phys. Rev. D}\ }\textbf {\bibinfo {volume}
  {109}},\ \bibinfo {pages} {L031104} (\bibinfo {year} {2024}{\natexlab{a}})},\
  \Eprint {http://arxiv.org/abs/2305.04982} {arXiv:2305.04982 [hep-ph]}
  \BibitemShut {NoStop}%
\bibitem [{\citenamefont {Afik}\ and\ \citenamefont
  {de~Nova}(2021)}]{Afik:2020onf}%
  \BibitemOpen
  \bibfield  {author} {\bibinfo {author} {\bibfnamefont {Y.}~\bibnamefont
  {Afik}}\ and\ \bibinfo {author} {\bibfnamefont {J.~R.~M.}\ \bibnamefont
  {de~Nova}},\ }\href {\doibase 10.1140/epjp/s13360-021-01902-1} {\bibfield
  {journal} {\bibinfo  {journal} {Eur. Phys. J. Plus}\ }\textbf {\bibinfo
  {volume} {136}},\ \bibinfo {pages} {907} (\bibinfo {year} {2021})},\ \Eprint
  {http://arxiv.org/abs/2003.02280} {arXiv:2003.02280 [quant-ph]} \BibitemShut
  {NoStop}%
\bibitem [{\citenamefont {Aad}\ \emph {et~al.}(2024)\citenamefont {Aad} \emph
  {et~al.}}]{ATLAS:2023fsd}%
  \BibitemOpen
  \bibfield  {author} {\bibinfo {author} {\bibfnamefont {G.}~\bibnamefont
  {Aad}} \emph {et~al.} (\bibinfo {collaboration} {ATLAS}),\ }\href {\doibase
  10.1038/s41586-024-07824-z} {\bibfield  {journal} {\bibinfo  {journal}
  {Nature}\ }\textbf {\bibinfo {volume} {633}},\ \bibinfo {pages} {542}
  (\bibinfo {year} {2024})},\ \Eprint {http://arxiv.org/abs/2311.07288}
  {arXiv:2311.07288 [hep-ex]} \BibitemShut {NoStop}%
\bibitem [{\citenamefont {Hayrapetyan}\ \emph {et~al.}(2024)\citenamefont
  {Hayrapetyan} \emph {et~al.}}]{CMS:2024pts}%
  \BibitemOpen
  \bibfield  {author} {\bibinfo {author} {\bibfnamefont {A.}~\bibnamefont
  {Hayrapetyan}} \emph {et~al.} (\bibinfo {collaboration} {CMS}),\ }\href
  {\doibase 10.1088/1361-6633/ad7e4d} {\bibfield  {journal} {\bibinfo
  {journal} {Rept. Prog. Phys.}\ }\textbf {\bibinfo {volume} {87}},\ \bibinfo
  {pages} {117801} (\bibinfo {year} {2024})},\ \Eprint
  {http://arxiv.org/abs/2406.03976} {arXiv:2406.03976 [hep-ex]} \BibitemShut
  {NoStop}%
\bibitem [{\citenamefont {Fabbrichesi}\ \emph {et~al.}(2021)\citenamefont
  {Fabbrichesi}, \citenamefont {Floreanini},\ and\ \citenamefont
  {Panizzo}}]{Fabbrichesi:2021npl}%
  \BibitemOpen
  \bibfield  {author} {\bibinfo {author} {\bibfnamefont {M.}~\bibnamefont
  {Fabbrichesi}}, \bibinfo {author} {\bibfnamefont {R.}~\bibnamefont
  {Floreanini}}, \ and\ \bibinfo {author} {\bibfnamefont {G.}~\bibnamefont
  {Panizzo}},\ }\href {\doibase 10.1103/PhysRevLett.127.161801} {\bibfield
  {journal} {\bibinfo  {journal} {Phys. Rev. Lett.}\ }\textbf {\bibinfo
  {volume} {127}},\ \bibinfo {pages} {161801} (\bibinfo {year} {2021})},\
  \Eprint {http://arxiv.org/abs/2102.11883} {arXiv:2102.11883 [hep-ph]}
  \BibitemShut {NoStop}%
\bibitem [{\citenamefont {Aoude}\ \emph {et~al.}(2022)\citenamefont {Aoude},
  \citenamefont {Madge}, \citenamefont {Maltoni},\ and\ \citenamefont
  {Mantani}}]{Aoude:2022imd}%
  \BibitemOpen
  \bibfield  {author} {\bibinfo {author} {\bibfnamefont {R.}~\bibnamefont
  {Aoude}}, \bibinfo {author} {\bibfnamefont {E.}~\bibnamefont {Madge}},
  \bibinfo {author} {\bibfnamefont {F.}~\bibnamefont {Maltoni}}, \ and\
  \bibinfo {author} {\bibfnamefont {L.}~\bibnamefont {Mantani}},\ }\href
  {\doibase 10.1103/PhysRevD.106.055007} {\bibfield  {journal} {\bibinfo
  {journal} {Phys. Rev. D}\ }\textbf {\bibinfo {volume} {106}},\ \bibinfo
  {pages} {055007} (\bibinfo {year} {2022})},\ \Eprint
  {http://arxiv.org/abs/2203.05619} {arXiv:2203.05619 [hep-ph]} \BibitemShut
  {NoStop}%
\bibitem [{\citenamefont {Afik}\ and\ \citenamefont
  {de~Nova}(2023)}]{Afik:2022dgh}%
  \BibitemOpen
  \bibfield  {author} {\bibinfo {author} {\bibfnamefont {Y.}~\bibnamefont
  {Afik}}\ and\ \bibinfo {author} {\bibfnamefont {J.~R.~M.}\ \bibnamefont
  {de~Nova}},\ }\href {\doibase 10.1103/PhysRevLett.130.221801} {\bibfield
  {journal} {\bibinfo  {journal} {Phys. Rev. Lett.}\ }\textbf {\bibinfo
  {volume} {130}},\ \bibinfo {pages} {221801} (\bibinfo {year} {2023})},\
  \Eprint {http://arxiv.org/abs/2209.03969} {arXiv:2209.03969 [quant-ph]}
  \BibitemShut {NoStop}%
\bibitem [{\citenamefont {Fabbrichesi}\ \emph
  {et~al.}(2023{\natexlab{a}})\citenamefont {Fabbrichesi}, \citenamefont
  {Floreanini},\ and\ \citenamefont {Gabrielli}}]{Fabbrichesi:2022ovb}%
  \BibitemOpen
  \bibfield  {author} {\bibinfo {author} {\bibfnamefont {M.}~\bibnamefont
  {Fabbrichesi}}, \bibinfo {author} {\bibfnamefont {R.}~\bibnamefont
  {Floreanini}}, \ and\ \bibinfo {author} {\bibfnamefont {E.}~\bibnamefont
  {Gabrielli}},\ }\href {\doibase 10.1140/epjc/s10052-023-11307-2} {\bibfield
  {journal} {\bibinfo  {journal} {Eur. Phys. J. C}\ }\textbf {\bibinfo {volume}
  {83}},\ \bibinfo {pages} {162} (\bibinfo {year} {2023}{\natexlab{a}})},\
  \Eprint {http://arxiv.org/abs/2208.11723} {arXiv:2208.11723 [hep-ph]}
  \BibitemShut {NoStop}%
\bibitem [{\citenamefont {Ehat{\"a}ht}\ \emph {et~al.}(2024)\citenamefont
  {Ehat{\"a}ht}, \citenamefont {Fabbrichesi}, \citenamefont {Marzola},\ and\
  \citenamefont {Veelken}}]{Ehataht:2023zzt}%
  \BibitemOpen
  \bibfield  {author} {\bibinfo {author} {\bibfnamefont {K.}~\bibnamefont
  {Ehat{\"a}ht}}, \bibinfo {author} {\bibfnamefont {M.}~\bibnamefont
  {Fabbrichesi}}, \bibinfo {author} {\bibfnamefont {L.}~\bibnamefont
  {Marzola}}, \ and\ \bibinfo {author} {\bibfnamefont {C.}~\bibnamefont
  {Veelken}},\ }\href {\doibase 10.1103/PhysRevD.109.032005} {\bibfield
  {journal} {\bibinfo  {journal} {Phys. Rev. D}\ }\textbf {\bibinfo {volume}
  {109}},\ \bibinfo {pages} {032005} (\bibinfo {year} {2024})},\ \Eprint
  {http://arxiv.org/abs/2311.17555} {arXiv:2311.17555 [hep-ph]} \BibitemShut
  {NoStop}%
\bibitem [{\citenamefont {Barr}(2022)}]{Barr:2021zcp}%
  \BibitemOpen
  \bibfield  {author} {\bibinfo {author} {\bibfnamefont {A.~J.}\ \bibnamefont
  {Barr}},\ }\href {\doibase 10.1016/j.physletb.2021.136866} {\bibfield
  {journal} {\bibinfo  {journal} {Phys. Lett. B}\ }\textbf {\bibinfo {volume}
  {825}},\ \bibinfo {pages} {136866} (\bibinfo {year} {2022})},\ \Eprint
  {http://arxiv.org/abs/2106.01377} {arXiv:2106.01377 [hep-ph]} \BibitemShut
  {NoStop}%
\bibitem [{\citenamefont {Barr}\ \emph {et~al.}(2023)\citenamefont {Barr},
  \citenamefont {Caban},\ and\ \citenamefont
  {Rembieli{\'n}ski}}]{Barr:2022wyq}%
  \BibitemOpen
  \bibfield  {author} {\bibinfo {author} {\bibfnamefont {A.~J.}\ \bibnamefont
  {Barr}}, \bibinfo {author} {\bibfnamefont {P.}~\bibnamefont {Caban}}, \ and\
  \bibinfo {author} {\bibfnamefont {J.}~\bibnamefont {Rembieli{\'n}ski}},\
  }\href {\doibase 10.22331/q-2023-07-27-1070} {\bibfield  {journal} {\bibinfo
  {journal} {Quantum}\ }\textbf {\bibinfo {volume} {7}},\ \bibinfo {pages}
  {1070} (\bibinfo {year} {2023})},\ \Eprint {http://arxiv.org/abs/2204.11063}
  {arXiv:2204.11063 [quant-ph]} \BibitemShut {NoStop}%
\bibitem [{\citenamefont {Aguilar-Saavedra}\ \emph {et~al.}(2023)\citenamefont
  {Aguilar-Saavedra}, \citenamefont {Bernal}, \citenamefont {Casas},\ and\
  \citenamefont {Moreno}}]{Aguilar-Saavedra:2022wam}%
  \BibitemOpen
  \bibfield  {author} {\bibinfo {author} {\bibfnamefont {J.~A.}\ \bibnamefont
  {Aguilar-Saavedra}}, \bibinfo {author} {\bibfnamefont {A.}~\bibnamefont
  {Bernal}}, \bibinfo {author} {\bibfnamefont {J.~A.}\ \bibnamefont {Casas}}, \
  and\ \bibinfo {author} {\bibfnamefont {J.~M.}\ \bibnamefont {Moreno}},\
  }\href {\doibase 10.1103/PhysRevD.107.016012} {\bibfield  {journal} {\bibinfo
   {journal} {Phys. Rev. D}\ }\textbf {\bibinfo {volume} {107}},\ \bibinfo
  {pages} {016012} (\bibinfo {year} {2023})},\ \Eprint
  {http://arxiv.org/abs/2209.13441} {arXiv:2209.13441 [hep-ph]} \BibitemShut
  {NoStop}%
\bibitem [{\citenamefont {Fabbrichesi}\ \emph
  {et~al.}(2023{\natexlab{b}})\citenamefont {Fabbrichesi}, \citenamefont
  {Floreanini}, \citenamefont {Gabrielli},\ and\ \citenamefont
  {Marzola}}]{Fabbrichesi:2023cev}%
  \BibitemOpen
  \bibfield  {author} {\bibinfo {author} {\bibfnamefont {M.}~\bibnamefont
  {Fabbrichesi}}, \bibinfo {author} {\bibfnamefont {R.}~\bibnamefont
  {Floreanini}}, \bibinfo {author} {\bibfnamefont {E.}~\bibnamefont
  {Gabrielli}}, \ and\ \bibinfo {author} {\bibfnamefont {L.}~\bibnamefont
  {Marzola}},\ }\href {\doibase 10.1140/epjc/s10052-023-11935-8} {\bibfield
  {journal} {\bibinfo  {journal} {Eur. Phys. J. C}\ }\textbf {\bibinfo {volume}
  {83}},\ \bibinfo {pages} {823} (\bibinfo {year} {2023}{\natexlab{b}})},\
  \Eprint {http://arxiv.org/abs/2302.00683} {arXiv:2302.00683 [hep-ph]}
  \BibitemShut {NoStop}%
\bibitem [{\citenamefont {Severi}\ \emph {et~al.}(2022)\citenamefont {Severi},
  \citenamefont {Boschi}, \citenamefont {Maltoni},\ and\ \citenamefont
  {Sioli}}]{Severi:2021cnj}%
  \BibitemOpen
  \bibfield  {author} {\bibinfo {author} {\bibfnamefont {C.}~\bibnamefont
  {Severi}}, \bibinfo {author} {\bibfnamefont {C.~D.~E.}\ \bibnamefont
  {Boschi}}, \bibinfo {author} {\bibfnamefont {F.}~\bibnamefont {Maltoni}}, \
  and\ \bibinfo {author} {\bibfnamefont {M.}~\bibnamefont {Sioli}},\ }\href
  {\doibase 10.1140/epjc/s10052-022-10245-9} {\bibfield  {journal} {\bibinfo
  {journal} {Eur. Phys. J. C}\ }\textbf {\bibinfo {volume} {82}},\ \bibinfo
  {pages} {285} (\bibinfo {year} {2022})},\ \Eprint
  {http://arxiv.org/abs/2110.10112} {arXiv:2110.10112 [hep-ph]} \BibitemShut
  {NoStop}%
\bibitem [{\citenamefont {Larkoski}(2022)}]{Larkoski:2022lmv}%
  \BibitemOpen
  \bibfield  {author} {\bibinfo {author} {\bibfnamefont {A.~J.}\ \bibnamefont
  {Larkoski}},\ }\href {\doibase 10.1103/PhysRevD.105.096012} {\bibfield
  {journal} {\bibinfo  {journal} {Phys. Rev. D}\ }\textbf {\bibinfo {volume}
  {105}},\ \bibinfo {pages} {096012} (\bibinfo {year} {2022})},\ \Eprint
  {http://arxiv.org/abs/2201.03159} {arXiv:2201.03159 [hep-ph]} \BibitemShut
  {NoStop}%
\bibitem [{\citenamefont {Aguilar-Saavedra}\ and\ \citenamefont
  {Casas}(2022)}]{Aguilar-Saavedra:2022uye}%
  \BibitemOpen
  \bibfield  {author} {\bibinfo {author} {\bibfnamefont {J.~A.}\ \bibnamefont
  {Aguilar-Saavedra}}\ and\ \bibinfo {author} {\bibfnamefont {J.~A.}\
  \bibnamefont {Casas}},\ }\href {\doibase 10.1140/epjc/s10052-022-10630-4}
  {\bibfield  {journal} {\bibinfo  {journal} {Eur. Phys. J. C}\ }\textbf
  {\bibinfo {volume} {82}},\ \bibinfo {pages} {666} (\bibinfo {year} {2022})},\
  \Eprint {http://arxiv.org/abs/2205.00542} {arXiv:2205.00542 [hep-ph]}
  \BibitemShut {NoStop}%
\bibitem [{\citenamefont {Afik}\ and\ \citenamefont
  {de~Nova}(2022)}]{Afik:2022kwm}%
  \BibitemOpen
  \bibfield  {author} {\bibinfo {author} {\bibfnamefont {Y.}~\bibnamefont
  {Afik}}\ and\ \bibinfo {author} {\bibfnamefont {J.~R.~M.}\ \bibnamefont
  {de~Nova}},\ }\href {\doibase 10.22331/q-2022-09-29-820} {\bibfield
  {journal} {\bibinfo  {journal} {Quantum}\ }\textbf {\bibinfo {volume} {6}},\
  \bibinfo {pages} {820} (\bibinfo {year} {2022})},\ \Eprint
  {http://arxiv.org/abs/2203.05582} {arXiv:2203.05582 [quant-ph]} \BibitemShut
  {NoStop}%
\bibitem [{\citenamefont {Gong}\ \emph {et~al.}(2022)\citenamefont {Gong},
  \citenamefont {Parida}, \citenamefont {Tu},\ and\ \citenamefont
  {Venugopalan}}]{Gong:2021bcp}%
  \BibitemOpen
  \bibfield  {author} {\bibinfo {author} {\bibfnamefont {W.}~\bibnamefont
  {Gong}}, \bibinfo {author} {\bibfnamefont {G.}~\bibnamefont {Parida}},
  \bibinfo {author} {\bibfnamefont {Z.}~\bibnamefont {Tu}}, \ and\ \bibinfo
  {author} {\bibfnamefont {R.}~\bibnamefont {Venugopalan}},\ }\href {\doibase
  10.1103/PhysRevD.106.L031501} {\bibfield  {journal} {\bibinfo  {journal}
  {Phys. Rev. D}\ }\textbf {\bibinfo {volume} {106}},\ \bibinfo {pages}
  {L031501} (\bibinfo {year} {2022})},\ \Eprint
  {http://arxiv.org/abs/2107.13007} {arXiv:2107.13007 [hep-ph]} \BibitemShut
  {NoStop}%
\bibitem [{\citenamefont
  {Aguilar-Saavedra}(2023{\natexlab{a}})}]{Aguilar-Saavedra:2023hss}%
  \BibitemOpen
  \bibfield  {author} {\bibinfo {author} {\bibfnamefont {J.~A.}\ \bibnamefont
  {Aguilar-Saavedra}},\ }\href {\doibase 10.1103/PhysRevD.108.076025}
  {\bibfield  {journal} {\bibinfo  {journal} {Phys. Rev. D}\ }\textbf {\bibinfo
  {volume} {108}},\ \bibinfo {pages} {076025} (\bibinfo {year}
  {2023}{\natexlab{a}})},\ \Eprint {http://arxiv.org/abs/2307.06991}
  {arXiv:2307.06991 [hep-ph]} \BibitemShut {NoStop}%
\bibitem [{\citenamefont {Aguilar-Saavedra}\ and\ \citenamefont
  {Casas}(2024)}]{Aguilar-Saavedra:2024fig}%
  \BibitemOpen
  \bibfield  {author} {\bibinfo {author} {\bibfnamefont {J.~A.}\ \bibnamefont
  {Aguilar-Saavedra}}\ and\ \bibinfo {author} {\bibfnamefont {J.~A.}\
  \bibnamefont {Casas}},\ }\href {\doibase 10.1103/PhysRevLett.133.111801}
  {\bibfield  {journal} {\bibinfo  {journal} {Phys. Rev. Lett.}\ }\textbf
  {\bibinfo {volume} {133}},\ \bibinfo {pages} {111801} (\bibinfo {year}
  {2024})},\ \Eprint {http://arxiv.org/abs/2401.06854} {arXiv:2401.06854
  [hep-ph]} \BibitemShut {NoStop}%
\bibitem [{\citenamefont {White}\ and\ \citenamefont
  {White}(2024)}]{White:2024nuc}%
  \BibitemOpen
  \bibfield  {author} {\bibinfo {author} {\bibfnamefont {C.~D.}\ \bibnamefont
  {White}}\ and\ \bibinfo {author} {\bibfnamefont {M.~J.}\ \bibnamefont
  {White}},\ }\href {\doibase 10.1103/PhysRevD.110.116016} {\bibfield
  {journal} {\bibinfo  {journal} {Phys. Rev. D}\ }\textbf {\bibinfo {volume}
  {110}},\ \bibinfo {pages} {116016} (\bibinfo {year} {2024})},\ \Eprint
  {http://arxiv.org/abs/2406.07321} {arXiv:2406.07321 [hep-ph]} \BibitemShut
  {NoStop}%
\bibitem [{\citenamefont {Han}\ \emph {et~al.}(2025{\natexlab{a}})\citenamefont
  {Han}, \citenamefont {Low}, \citenamefont {McGinnis},\ and\ \citenamefont
  {Su}}]{Han:2024ugl}%
  \BibitemOpen
  \bibfield  {author} {\bibinfo {author} {\bibfnamefont {T.}~\bibnamefont
  {Han}}, \bibinfo {author} {\bibfnamefont {M.}~\bibnamefont {Low}}, \bibinfo
  {author} {\bibfnamefont {N.}~\bibnamefont {McGinnis}}, \ and\ \bibinfo
  {author} {\bibfnamefont {S.}~\bibnamefont {Su}},\ }\href {\doibase
  10.1007/JHEP05(2025)081} {\bibfield  {journal} {\bibinfo  {journal} {JHEP}\
  }\textbf {\bibinfo {volume} {05}},\ \bibinfo {pages} {081} (\bibinfo {year}
  {2025}{\natexlab{a}})},\ \Eprint {http://arxiv.org/abs/2412.21158}
  {arXiv:2412.21158 [hep-ph]} \BibitemShut {NoStop}%
\bibitem [{\citenamefont {Fabbrichesi}\ \emph {et~al.}(2025)\citenamefont
  {Fabbrichesi}, \citenamefont {Low},\ and\ \citenamefont
  {Marzola}}]{Fabbrichesi:2025ywl}%
  \BibitemOpen
  \bibfield  {author} {\bibinfo {author} {\bibfnamefont {M.}~\bibnamefont
  {Fabbrichesi}}, \bibinfo {author} {\bibfnamefont {M.}~\bibnamefont {Low}}, \
  and\ \bibinfo {author} {\bibfnamefont {L.}~\bibnamefont {Marzola}},\ }\href
  {\doibase 10.1103/kdmh-3yb4} {\bibfield  {journal} {\bibinfo  {journal}
  {Phys. Rev. D}\ }\textbf {\bibinfo {volume} {112}},\ \bibinfo {pages}
  {013003} (\bibinfo {year} {2025})},\ \Eprint
  {http://arxiv.org/abs/2501.03311} {arXiv:2501.03311 [hep-ph]} \BibitemShut
  {NoStop}%
\bibitem [{\citenamefont {Ashby-Pickering}\ \emph {et~al.}(2023)\citenamefont
  {Ashby-Pickering}, \citenamefont {Barr},\ and\ \citenamefont
  {Wierzchucka}}]{Ashby-Pickering:2022umy}%
  \BibitemOpen
  \bibfield  {author} {\bibinfo {author} {\bibfnamefont {R.}~\bibnamefont
  {Ashby-Pickering}}, \bibinfo {author} {\bibfnamefont {A.~J.}\ \bibnamefont
  {Barr}}, \ and\ \bibinfo {author} {\bibfnamefont {A.}~\bibnamefont
  {Wierzchucka}},\ }\href {\doibase 10.1007/JHEP05(2023)020} {\bibfield
  {journal} {\bibinfo  {journal} {JHEP}\ }\textbf {\bibinfo {volume} {05}},\
  \bibinfo {pages} {020} (\bibinfo {year} {2023})},\ \Eprint
  {http://arxiv.org/abs/2209.13990} {arXiv:2209.13990 [quant-ph]} \BibitemShut
  {NoStop}%
\bibitem [{\citenamefont
  {Aguilar-Saavedra}(2023{\natexlab{b}})}]{Aguilar-Saavedra:2022mpg}%
  \BibitemOpen
  \bibfield  {author} {\bibinfo {author} {\bibfnamefont {J.~A.}\ \bibnamefont
  {Aguilar-Saavedra}},\ }\href {\doibase 10.1103/PhysRevD.107.076016}
  {\bibfield  {journal} {\bibinfo  {journal} {Phys. Rev. D}\ }\textbf {\bibinfo
  {volume} {107}},\ \bibinfo {pages} {076016} (\bibinfo {year}
  {2023}{\natexlab{b}})},\ \Eprint {http://arxiv.org/abs/2209.14033}
  {arXiv:2209.14033 [hep-ph]} \BibitemShut {NoStop}%
\bibitem [{\citenamefont {Altakach}\ \emph {et~al.}(2023)\citenamefont
  {Altakach}, \citenamefont {Lamba}, \citenamefont {Maltoni}, \citenamefont
  {Mawatari},\ and\ \citenamefont {Sakurai}}]{Altakach:2022ywa}%
  \BibitemOpen
  \bibfield  {author} {\bibinfo {author} {\bibfnamefont {M.~M.}\ \bibnamefont
  {Altakach}}, \bibinfo {author} {\bibfnamefont {P.}~\bibnamefont {Lamba}},
  \bibinfo {author} {\bibfnamefont {F.}~\bibnamefont {Maltoni}}, \bibinfo
  {author} {\bibfnamefont {K.}~\bibnamefont {Mawatari}}, \ and\ \bibinfo
  {author} {\bibfnamefont {K.}~\bibnamefont {Sakurai}},\ }\href {\doibase
  10.1103/PhysRevD.107.093002} {\bibfield  {journal} {\bibinfo  {journal}
  {Phys. Rev. D}\ }\textbf {\bibinfo {volume} {107}},\ \bibinfo {pages}
  {093002} (\bibinfo {year} {2023})},\ \Eprint
  {http://arxiv.org/abs/2211.10513} {arXiv:2211.10513 [hep-ph]} \BibitemShut
  {NoStop}%
\bibitem [{\citenamefont {Aoude}\ \emph {et~al.}(2023)\citenamefont {Aoude},
  \citenamefont {Madge}, \citenamefont {Maltoni},\ and\ \citenamefont
  {Mantani}}]{Aoude:2023hxv}%
  \BibitemOpen
  \bibfield  {author} {\bibinfo {author} {\bibfnamefont {R.}~\bibnamefont
  {Aoude}}, \bibinfo {author} {\bibfnamefont {E.}~\bibnamefont {Madge}},
  \bibinfo {author} {\bibfnamefont {F.}~\bibnamefont {Maltoni}}, \ and\
  \bibinfo {author} {\bibfnamefont {L.}~\bibnamefont {Mantani}},\ }\href
  {\doibase 10.1007/JHEP12(2023)017} {\bibfield  {journal} {\bibinfo  {journal}
  {JHEP}\ }\textbf {\bibinfo {volume} {12}},\ \bibinfo {pages} {017} (\bibinfo
  {year} {2023})},\ \Eprint {http://arxiv.org/abs/2307.09675} {arXiv:2307.09675
  [hep-ph]} \BibitemShut {NoStop}%
\bibitem [{\citenamefont {Morales}(2023)}]{Morales:2023gow}%
  \BibitemOpen
  \bibfield  {author} {\bibinfo {author} {\bibfnamefont {R.~A.}\ \bibnamefont
  {Morales}},\ }\href {\doibase 10.1140/epjp/s13360-023-04784-7} {\bibfield
  {journal} {\bibinfo  {journal} {Eur. Phys. J. Plus}\ }\textbf {\bibinfo
  {volume} {138}},\ \bibinfo {pages} {1157} (\bibinfo {year} {2023})},\ \Eprint
  {http://arxiv.org/abs/2306.17247} {arXiv:2306.17247 [hep-ph]} \BibitemShut
  {NoStop}%
\bibitem [{\citenamefont {Bernal}\ \emph {et~al.}(2023)\citenamefont {Bernal},
  \citenamefont {Caban},\ and\ \citenamefont
  {Rembieli{\'n}ski}}]{Bernal:2023ruk}%
  \BibitemOpen
  \bibfield  {author} {\bibinfo {author} {\bibfnamefont {A.}~\bibnamefont
  {Bernal}}, \bibinfo {author} {\bibfnamefont {P.}~\bibnamefont {Caban}}, \
  and\ \bibinfo {author} {\bibfnamefont {J.}~\bibnamefont {Rembieli{\'n}ski}},\
  }\href {\doibase 10.1140/epjc/s10052-023-12216-0} {\bibfield  {journal}
  {\bibinfo  {journal} {Eur. Phys. J. C}\ }\textbf {\bibinfo {volume} {83}},\
  \bibinfo {pages} {1050} (\bibinfo {year} {2023})},\ \Eprint
  {http://arxiv.org/abs/2307.13496} {arXiv:2307.13496 [hep-ph]} \BibitemShut
  {NoStop}%
\bibitem [{\citenamefont {Bi}\ \emph {et~al.}(2024)\citenamefont {Bi},
  \citenamefont {Cao}, \citenamefont {Cheng},\ and\ \citenamefont
  {Zhang}}]{Bi:2023uop}%
  \BibitemOpen
  \bibfield  {author} {\bibinfo {author} {\bibfnamefont {Q.}~\bibnamefont
  {Bi}}, \bibinfo {author} {\bibfnamefont {Q.-H.}\ \bibnamefont {Cao}},
  \bibinfo {author} {\bibfnamefont {K.}~\bibnamefont {Cheng}}, \ and\ \bibinfo
  {author} {\bibfnamefont {H.}~\bibnamefont {Zhang}},\ }\href {\doibase
  10.1103/PhysRevD.109.036022} {\bibfield  {journal} {\bibinfo  {journal}
  {Phys. Rev. D}\ }\textbf {\bibinfo {volume} {109}},\ \bibinfo {pages}
  {036022} (\bibinfo {year} {2024})},\ \Eprint
  {http://arxiv.org/abs/2307.14895} {arXiv:2307.14895 [hep-ph]} \BibitemShut
  {NoStop}%
\bibitem [{\citenamefont {Dong}\ \emph {et~al.}(2024)\citenamefont {Dong},
  \citenamefont {Gon{\c{c}}alves}, \citenamefont {Kong},\ and\ \citenamefont
  {Navarro}}]{Dong:2023xiw}%
  \BibitemOpen
  \bibfield  {author} {\bibinfo {author} {\bibfnamefont {Z.}~\bibnamefont
  {Dong}}, \bibinfo {author} {\bibfnamefont {D.}~\bibnamefont
  {Gon{\c{c}}alves}}, \bibinfo {author} {\bibfnamefont {K.}~\bibnamefont
  {Kong}}, \ and\ \bibinfo {author} {\bibfnamefont {A.}~\bibnamefont
  {Navarro}},\ }\href {\doibase 10.1103/PhysRevD.109.115023} {\bibfield
  {journal} {\bibinfo  {journal} {Phys. Rev. D}\ }\textbf {\bibinfo {volume}
  {109}},\ \bibinfo {pages} {115023} (\bibinfo {year} {2024})},\ \Eprint
  {http://arxiv.org/abs/2305.07075} {arXiv:2305.07075 [hep-ph]} \BibitemShut
  {NoStop}%
\bibitem [{\citenamefont {Ma}\ and\ \citenamefont {Li}(2024)}]{Ma:2023yvd}%
  \BibitemOpen
  \bibfield  {author} {\bibinfo {author} {\bibfnamefont {K.}~\bibnamefont
  {Ma}}\ and\ \bibinfo {author} {\bibfnamefont {T.}~\bibnamefont {Li}},\ }\href
  {\doibase 10.1088/1674-1137/ad62d8} {\bibfield  {journal} {\bibinfo
  {journal} {Chin. Phys. C}\ }\textbf {\bibinfo {volume} {48}},\ \bibinfo
  {pages} {103105} (\bibinfo {year} {2024})},\ \Eprint
  {http://arxiv.org/abs/2309.08103} {arXiv:2309.08103 [hep-ph]} \BibitemShut
  {NoStop}%
\bibitem [{\citenamefont {Sakurai}\ and\ \citenamefont
  {Spannowsky}(2024)}]{Sakurai:2023nsc}%
  \BibitemOpen
  \bibfield  {author} {\bibinfo {author} {\bibfnamefont {K.}~\bibnamefont
  {Sakurai}}\ and\ \bibinfo {author} {\bibfnamefont {M.}~\bibnamefont
  {Spannowsky}},\ }\href {\doibase 10.1103/PhysRevLett.132.151602} {\bibfield
  {journal} {\bibinfo  {journal} {Phys. Rev. Lett.}\ }\textbf {\bibinfo
  {volume} {132}},\ \bibinfo {pages} {151602} (\bibinfo {year} {2024})},\
  \Eprint {http://arxiv.org/abs/2310.01477} {arXiv:2310.01477 [quant-ph]}
  \BibitemShut {NoStop}%
\bibitem [{\citenamefont {Bernal}(2024)}]{Bernal:2023jba}%
  \BibitemOpen
  \bibfield  {author} {\bibinfo {author} {\bibfnamefont {A.}~\bibnamefont
  {Bernal}},\ }\href {\doibase 10.1103/PhysRevD.109.116007} {\bibfield
  {journal} {\bibinfo  {journal} {Phys. Rev. D}\ }\textbf {\bibinfo {volume}
  {109}},\ \bibinfo {pages} {116007} (\bibinfo {year} {2024})},\ \Eprint
  {http://arxiv.org/abs/2310.10838} {arXiv:2310.10838 [hep-ph]} \BibitemShut
  {NoStop}%
\bibitem [{\citenamefont {Han}\ \emph {et~al.}(2024)\citenamefont {Han},
  \citenamefont {Low},\ and\ \citenamefont {Wu}}]{Han:2023fci}%
  \BibitemOpen
  \bibfield  {author} {\bibinfo {author} {\bibfnamefont {T.}~\bibnamefont
  {Han}}, \bibinfo {author} {\bibfnamefont {M.}~\bibnamefont {Low}}, \ and\
  \bibinfo {author} {\bibfnamefont {T.~A.}\ \bibnamefont {Wu}},\ }\href
  {\doibase 10.1007/JHEP07(2024)192} {\bibfield  {journal} {\bibinfo  {journal}
  {JHEP}\ }\textbf {\bibinfo {volume} {07}},\ \bibinfo {pages} {192} (\bibinfo
  {year} {2024})},\ \Eprint {http://arxiv.org/abs/2310.17696} {arXiv:2310.17696
  [hep-ph]} \BibitemShut {NoStop}%
\bibitem [{\citenamefont {Cheng}\ \emph {et~al.}(2024)\citenamefont {Cheng},
  \citenamefont {Han},\ and\ \citenamefont {Low}}]{Cheng:2023qmz}%
  \BibitemOpen
  \bibfield  {author} {\bibinfo {author} {\bibfnamefont {K.}~\bibnamefont
  {Cheng}}, \bibinfo {author} {\bibfnamefont {T.}~\bibnamefont {Han}}, \ and\
  \bibinfo {author} {\bibfnamefont {M.}~\bibnamefont {Low}},\ }\href {\doibase
  10.1103/PhysRevD.109.116005} {\bibfield  {journal} {\bibinfo  {journal}
  {Phys. Rev. D}\ }\textbf {\bibinfo {volume} {109}},\ \bibinfo {pages}
  {116005} (\bibinfo {year} {2024})},\ \Eprint
  {http://arxiv.org/abs/2311.09166} {arXiv:2311.09166 [hep-ph]} \BibitemShut
  {NoStop}%
\bibitem [{\citenamefont
  {Aguilar-Saavedra}(2024{\natexlab{a}})}]{Aguilar-Saavedra:2024hwd}%
  \BibitemOpen
  \bibfield  {author} {\bibinfo {author} {\bibfnamefont {J.~A.}\ \bibnamefont
  {Aguilar-Saavedra}},\ }\href {\doibase 10.1103/PhysRevD.109.096027}
  {\bibfield  {journal} {\bibinfo  {journal} {Phys. Rev. D}\ }\textbf {\bibinfo
  {volume} {109}},\ \bibinfo {pages} {096027} (\bibinfo {year}
  {2024}{\natexlab{a}})},\ \Eprint {http://arxiv.org/abs/2401.10988}
  {arXiv:2401.10988 [hep-ph]} \BibitemShut {NoStop}%
\bibitem [{\citenamefont
  {Aguilar-Saavedra}(2024{\natexlab{b}})}]{Aguilar-Saavedra:2024vpd}%
  \BibitemOpen
  \bibfield  {author} {\bibinfo {author} {\bibfnamefont {J.~A.}\ \bibnamefont
  {Aguilar-Saavedra}},\ }\href {\doibase 10.1016/j.physletb.2024.138849}
  {\bibfield  {journal} {\bibinfo  {journal} {Phys. Lett. B}\ }\textbf
  {\bibinfo {volume} {855}},\ \bibinfo {pages} {138849} (\bibinfo {year}
  {2024}{\natexlab{b}})},\ \Eprint {http://arxiv.org/abs/2402.14725}
  {arXiv:2402.14725 [hep-ph]} \BibitemShut {NoStop}%
\bibitem [{\citenamefont
  {Aguilar-Saavedra}(2024{\natexlab{c}})}]{Aguilar-Saavedra:2024whi}%
  \BibitemOpen
  \bibfield  {author} {\bibinfo {author} {\bibfnamefont {J.~A.}\ \bibnamefont
  {Aguilar-Saavedra}},\ }\href {\doibase 10.1103/PhysRevD.109.113004}
  {\bibfield  {journal} {\bibinfo  {journal} {Phys. Rev. D}\ }\textbf {\bibinfo
  {volume} {109}},\ \bibinfo {pages} {113004} (\bibinfo {year}
  {2024}{\natexlab{c}})},\ \Eprint {http://arxiv.org/abs/2403.13942}
  {arXiv:2403.13942 [hep-ph]} \BibitemShut {NoStop}%
\bibitem [{\citenamefont {Duch}\ \emph {et~al.}(2025)\citenamefont {Duch},
  \citenamefont {Strumia},\ and\ \citenamefont {Titov}}]{Duch:2024pwm}%
  \BibitemOpen
  \bibfield  {author} {\bibinfo {author} {\bibfnamefont {M.}~\bibnamefont
  {Duch}}, \bibinfo {author} {\bibfnamefont {A.}~\bibnamefont {Strumia}}, \
  and\ \bibinfo {author} {\bibfnamefont {A.}~\bibnamefont {Titov}},\ }\href
  {\doibase 10.1140/epjc/s10052-025-13836-4} {\bibfield  {journal} {\bibinfo
  {journal} {Eur. Phys. J. C}\ }\textbf {\bibinfo {volume} {85}},\ \bibinfo
  {pages} {151} (\bibinfo {year} {2025})},\ \Eprint
  {http://arxiv.org/abs/2403.14757} {arXiv:2403.14757 [hep-ph]} \BibitemShut
  {NoStop}%
\bibitem [{\citenamefont {Morales}(2024)}]{Morales:2024jhj}%
  \BibitemOpen
  \bibfield  {author} {\bibinfo {author} {\bibfnamefont {R.~A.}\ \bibnamefont
  {Morales}},\ }\href {\doibase 10.1140/epjc/s10052-024-12921-4} {\bibfield
  {journal} {\bibinfo  {journal} {Eur. Phys. J. C}\ }\textbf {\bibinfo {volume}
  {84}},\ \bibinfo {pages} {581} (\bibinfo {year} {2024})},\ \Eprint
  {http://arxiv.org/abs/2403.18023} {arXiv:2403.18023 [hep-ph]} \BibitemShut
  {NoStop}%
\bibitem [{\citenamefont {Subba}\ and\ \citenamefont
  {Rahaman}(2024)}]{Subba:2024mnl}%
  \BibitemOpen
  \bibfield  {author} {\bibinfo {author} {\bibfnamefont {A.}~\bibnamefont
  {Subba}}\ and\ \bibinfo {author} {\bibfnamefont {R.}~\bibnamefont
  {Rahaman}},\ }\href@noop {} {\  (\bibinfo {year} {2024})},\ \Eprint
  {http://arxiv.org/abs/2404.03292} {arXiv:2404.03292 [hep-ph]} \BibitemShut
  {NoStop}%
\bibitem [{\citenamefont {Maltoni}\ \emph {et~al.}(2024)\citenamefont
  {Maltoni}, \citenamefont {Severi}, \citenamefont {Tentori},\ and\
  \citenamefont {Vryonidou}}]{Maltoni:2024csn}%
  \BibitemOpen
  \bibfield  {author} {\bibinfo {author} {\bibfnamefont {F.}~\bibnamefont
  {Maltoni}}, \bibinfo {author} {\bibfnamefont {C.}~\bibnamefont {Severi}},
  \bibinfo {author} {\bibfnamefont {S.}~\bibnamefont {Tentori}}, \ and\
  \bibinfo {author} {\bibfnamefont {E.}~\bibnamefont {Vryonidou}},\ }\href
  {\doibase 10.1007/JHEP09(2024)001} {\bibfield  {journal} {\bibinfo  {journal}
  {JHEP}\ }\textbf {\bibinfo {volume} {09}},\ \bibinfo {pages} {001} (\bibinfo
  {year} {2024})},\ \Eprint {http://arxiv.org/abs/2404.08049} {arXiv:2404.08049
  [hep-ph]} \BibitemShut {NoStop}%
\bibitem [{\citenamefont {Afik}\ \emph {et~al.}(2025)\citenamefont {Afik},
  \citenamefont {Kats}, \citenamefont {de~Nova}, \citenamefont {Soffer},\ and\
  \citenamefont {Uzan}}]{Afik:2025grr}%
  \BibitemOpen
  \bibfield  {author} {\bibinfo {author} {\bibfnamefont {Y.}~\bibnamefont
  {Afik}}, \bibinfo {author} {\bibfnamefont {Y.}~\bibnamefont {Kats}}, \bibinfo
  {author} {\bibfnamefont {J.~R.~M.}\ \bibnamefont {de~Nova}}, \bibinfo
  {author} {\bibfnamefont {A.}~\bibnamefont {Soffer}}, \ and\ \bibinfo {author}
  {\bibfnamefont {D.}~\bibnamefont {Uzan}},\ }\href {\doibase
  10.1103/fhkc-kfhr} {\bibfield  {journal} {\bibinfo  {journal} {Phys. Rev. D}\
  }\textbf {\bibinfo {volume} {111}},\ \bibinfo {pages} {L111902} (\bibinfo
  {year} {2025})},\ \Eprint {http://arxiv.org/abs/2406.04402} {arXiv:2406.04402
  [hep-ph]} \BibitemShut {NoStop}%
\bibitem [{\citenamefont {Wu}\ \emph {et~al.}(2024)\citenamefont {Wu},
  \citenamefont {Qian}, \citenamefont {Wang},\ and\ \citenamefont
  {Zhou}}]{Wu:2024asu}%
  \BibitemOpen
  \bibfield  {author} {\bibinfo {author} {\bibfnamefont {S.}~\bibnamefont
  {Wu}}, \bibinfo {author} {\bibfnamefont {C.}~\bibnamefont {Qian}}, \bibinfo
  {author} {\bibfnamefont {Q.}~\bibnamefont {Wang}}, \ and\ \bibinfo {author}
  {\bibfnamefont {X.-R.}\ \bibnamefont {Zhou}},\ }\href {\doibase
  10.1103/PhysRevD.110.054012} {\bibfield  {journal} {\bibinfo  {journal}
  {Phys. Rev. D}\ }\textbf {\bibinfo {volume} {110}},\ \bibinfo {pages}
  {054012} (\bibinfo {year} {2024})},\ \Eprint
  {http://arxiv.org/abs/2406.16298} {arXiv:2406.16298 [hep-ph]} \BibitemShut
  {NoStop}%
\bibitem [{\citenamefont {Cheng}\ \emph
  {et~al.}(2025{\natexlab{a}})\citenamefont {Cheng}, \citenamefont {Han},\ and\
  \citenamefont {Low}}]{Cheng:2024btk}%
  \BibitemOpen
  \bibfield  {author} {\bibinfo {author} {\bibfnamefont {K.}~\bibnamefont
  {Cheng}}, \bibinfo {author} {\bibfnamefont {T.}~\bibnamefont {Han}}, \ and\
  \bibinfo {author} {\bibfnamefont {M.}~\bibnamefont {Low}},\ }\href {\doibase
  10.1103/PhysRevD.111.033004} {\bibfield  {journal} {\bibinfo  {journal}
  {Phys. Rev. D}\ }\textbf {\bibinfo {volume} {111}},\ \bibinfo {pages}
  {033004} (\bibinfo {year} {2025}{\natexlab{a}})},\ \Eprint
  {http://arxiv.org/abs/2407.01672} {arXiv:2407.01672 [hep-ph]} \BibitemShut
  {NoStop}%
\bibitem [{\citenamefont {Gabrielli}\ and\ \citenamefont
  {Marzola}(2024)}]{Gabrielli:2024kbz}%
  \BibitemOpen
  \bibfield  {author} {\bibinfo {author} {\bibfnamefont {E.}~\bibnamefont
  {Gabrielli}}\ and\ \bibinfo {author} {\bibfnamefont {L.}~\bibnamefont
  {Marzola}},\ }\href {\doibase 10.3390/sym16081036} {\bibfield  {journal}
  {\bibinfo  {journal} {Symmetry}\ }\textbf {\bibinfo {volume} {16}},\ \bibinfo
  {pages} {1036} (\bibinfo {year} {2024})},\ \Eprint
  {http://arxiv.org/abs/2408.05010} {arXiv:2408.05010 [hep-ph]} \BibitemShut
  {NoStop}%
\bibitem [{\citenamefont {Ruzi}\ \emph {et~al.}(2024)\citenamefont {Ruzi},
  \citenamefont {Wu}, \citenamefont {Ding}, \citenamefont {Qian}, \citenamefont
  {Levin},\ and\ \citenamefont {Li}}]{Ruzi:2024cbt}%
  \BibitemOpen
  \bibfield  {author} {\bibinfo {author} {\bibfnamefont {A.}~\bibnamefont
  {Ruzi}}, \bibinfo {author} {\bibfnamefont {Y.}~\bibnamefont {Wu}}, \bibinfo
  {author} {\bibfnamefont {R.}~\bibnamefont {Ding}}, \bibinfo {author}
  {\bibfnamefont {S.}~\bibnamefont {Qian}}, \bibinfo {author} {\bibfnamefont
  {A.~M.}\ \bibnamefont {Levin}}, \ and\ \bibinfo {author} {\bibfnamefont
  {Q.}~\bibnamefont {Li}},\ }\href {\doibase 10.1007/JHEP10(2024)211}
  {\bibfield  {journal} {\bibinfo  {journal} {JHEP}\ }\textbf {\bibinfo
  {volume} {10}},\ \bibinfo {pages} {211} (\bibinfo {year} {2024})},\ \Eprint
  {http://arxiv.org/abs/2408.05429} {arXiv:2408.05429 [hep-ph]} \BibitemShut
  {NoStop}%
\bibitem [{\citenamefont {Cheng}\ \emph
  {et~al.}(2025{\natexlab{b}})\citenamefont {Cheng}, \citenamefont {Han},\ and\
  \citenamefont {Low}}]{Cheng:2024rxi}%
  \BibitemOpen
  \bibfield  {author} {\bibinfo {author} {\bibfnamefont {K.}~\bibnamefont
  {Cheng}}, \bibinfo {author} {\bibfnamefont {T.}~\bibnamefont {Han}}, \ and\
  \bibinfo {author} {\bibfnamefont {M.}~\bibnamefont {Low}},\ }\href {\doibase
  10.1016/j.physletb.2025.139675} {\bibfield  {journal} {\bibinfo  {journal}
  {Phys. Lett. B}\ }\textbf {\bibinfo {volume} {868}},\ \bibinfo {pages}
  {139675} (\bibinfo {year} {2025}{\natexlab{b}})},\ \Eprint
  {http://arxiv.org/abs/2410.08303} {arXiv:2410.08303 [hep-ph]} \BibitemShut
  {NoStop}%
\bibitem [{\citenamefont {Wu}\ \emph {et~al.}(2025)\citenamefont {Wu},
  \citenamefont {Jiang}, \citenamefont {Ruzi}, \citenamefont {Ban},
  \citenamefont {Yan},\ and\ \citenamefont {Li}}]{Wu:2024ovc}%
  \BibitemOpen
  \bibfield  {author} {\bibinfo {author} {\bibfnamefont {Y.}~\bibnamefont
  {Wu}}, \bibinfo {author} {\bibfnamefont {R.}~\bibnamefont {Jiang}}, \bibinfo
  {author} {\bibfnamefont {A.}~\bibnamefont {Ruzi}}, \bibinfo {author}
  {\bibfnamefont {Y.}~\bibnamefont {Ban}}, \bibinfo {author} {\bibfnamefont
  {X.}~\bibnamefont {Yan}}, \ and\ \bibinfo {author} {\bibfnamefont
  {Q.}~\bibnamefont {Li}},\ }\href {\doibase 10.1103/PhysRevD.111.036008}
  {\bibfield  {journal} {\bibinfo  {journal} {Phys. Rev. D}\ }\textbf {\bibinfo
  {volume} {111}},\ \bibinfo {pages} {036008} (\bibinfo {year} {2025})},\
  \Eprint {http://arxiv.org/abs/2410.17025} {arXiv:2410.17025 [hep-ph]}
  \BibitemShut {NoStop}%
\bibitem [{\citenamefont {Ruzi}\ \emph {et~al.}(2025)\citenamefont {Ruzi},
  \citenamefont {Gao}, \citenamefont {Li}, \citenamefont {Zhou}, \citenamefont
  {Chen}, \citenamefont {Zhang}, \citenamefont {Sun},\ and\ \citenamefont
  {Li}}]{Ruzi:2024iqu}%
  \BibitemOpen
  \bibfield  {author} {\bibinfo {author} {\bibfnamefont {A.}~\bibnamefont
  {Ruzi}}, \bibinfo {author} {\bibfnamefont {L.}~\bibnamefont {Gao}}, \bibinfo
  {author} {\bibfnamefont {Q.}~\bibnamefont {Li}}, \bibinfo {author}
  {\bibfnamefont {C.}~\bibnamefont {Zhou}}, \bibinfo {author} {\bibfnamefont
  {L.}~\bibnamefont {Chen}}, \bibinfo {author} {\bibfnamefont {X.}~\bibnamefont
  {Zhang}}, \bibinfo {author} {\bibfnamefont {Z.}~\bibnamefont {Sun}}, \ and\
  \bibinfo {author} {\bibfnamefont {Q.}~\bibnamefont {Li}},\ }\href {\doibase
  10.1088/1361-6471/ade733} {\bibfield  {journal} {\bibinfo  {journal} {J.
  Phys. G}\ }\textbf {\bibinfo {volume} {52}},\ \bibinfo {pages} {075002}
  (\bibinfo {year} {2025})},\ \Eprint {http://arxiv.org/abs/2411.12518}
  {arXiv:2411.12518 [hep-ph]} \BibitemShut {NoStop}%
\bibitem [{\citenamefont {Altomonte}\ \emph {et~al.}(2025)\citenamefont
  {Altomonte}, \citenamefont {Barr}, \citenamefont {Eckstein}, \citenamefont
  {Horodecki},\ and\ \citenamefont {Sakurai}}]{Altomonte:2024upf}%
  \BibitemOpen
  \bibfield  {author} {\bibinfo {author} {\bibfnamefont {C.}~\bibnamefont
  {Altomonte}}, \bibinfo {author} {\bibfnamefont {A.~J.}\ \bibnamefont {Barr}},
  \bibinfo {author} {\bibfnamefont {M.}~\bibnamefont {Eckstein}}, \bibinfo
  {author} {\bibfnamefont {P.}~\bibnamefont {Horodecki}}, \ and\ \bibinfo
  {author} {\bibfnamefont {K.}~\bibnamefont {Sakurai}},\ }\href {\doibase
  10.1088/2058-9565/ae0af1} {\bibfield  {journal} {\bibinfo  {journal} {Quantum
  Sci. Technol.}\ }\textbf {\bibinfo {volume} {10}},\ \bibinfo {pages} {045060}
  (\bibinfo {year} {2025})},\ \Eprint {http://arxiv.org/abs/2412.01892}
  {arXiv:2412.01892 [hep-ph]} \BibitemShut {NoStop}%
\bibitem [{\citenamefont {Fabbrichesi}\ \emph
  {et~al.}(2024{\natexlab{b}})\citenamefont {Fabbrichesi}, \citenamefont
  {Floreanini}, \citenamefont {Gabrielli},\ and\ \citenamefont
  {Marzola}}]{Fabbrichesi:2024rec}%
  \BibitemOpen
  \bibfield  {author} {\bibinfo {author} {\bibfnamefont {M.}~\bibnamefont
  {Fabbrichesi}}, \bibinfo {author} {\bibfnamefont {R.}~\bibnamefont
  {Floreanini}}, \bibinfo {author} {\bibfnamefont {E.}~\bibnamefont
  {Gabrielli}}, \ and\ \bibinfo {author} {\bibfnamefont {L.}~\bibnamefont
  {Marzola}},\ }\href {\doibase 10.1103/PhysRevD.110.053008} {\bibfield
  {journal} {\bibinfo  {journal} {Phys. Rev. D}\ }\textbf {\bibinfo {volume}
  {110}},\ \bibinfo {pages} {053008} (\bibinfo {year} {2024}{\natexlab{b}})},\
  \Eprint {http://arxiv.org/abs/2406.17772} {arXiv:2406.17772 [hep-ph]}
  \BibitemShut {NoStop}%
\bibitem [{\citenamefont {Cheng}\ and\ \citenamefont
  {Yan}(2025)}]{Cheng:2025cuv}%
  \BibitemOpen
  \bibfield  {author} {\bibinfo {author} {\bibfnamefont {K.}~\bibnamefont
  {Cheng}}\ and\ \bibinfo {author} {\bibfnamefont {B.}~\bibnamefont {Yan}},\
  }\href {\doibase 10.1103/gmqz-v4cl} {\bibfield  {journal} {\bibinfo
  {journal} {Phys. Rev. Lett.}\ }\textbf {\bibinfo {volume} {135}},\ \bibinfo
  {pages} {011902} (\bibinfo {year} {2025})},\ \Eprint
  {http://arxiv.org/abs/2501.03321} {arXiv:2501.03321 [hep-ph]} \BibitemShut
  {NoStop}%
\bibitem [{\citenamefont {Han}\ \emph {et~al.}(2025{\natexlab{b}})\citenamefont
  {Han}, \citenamefont {Low},\ and\ \citenamefont {Su}}]{Han:2025ewp}%
  \BibitemOpen
  \bibfield  {author} {\bibinfo {author} {\bibfnamefont {T.}~\bibnamefont
  {Han}}, \bibinfo {author} {\bibfnamefont {M.}~\bibnamefont {Low}}, \ and\
  \bibinfo {author} {\bibfnamefont {Y.}~\bibnamefont {Su}},\ }\href {\doibase
  10.1007/JHEP10(2025)217} {\bibfield  {journal} {\bibinfo  {journal} {JHEP}\
  }\textbf {\bibinfo {volume} {10}},\ \bibinfo {pages} {217} (\bibinfo {year}
  {2025}{\natexlab{b}})},\ \Eprint {http://arxiv.org/abs/2501.04801}
  {arXiv:2501.04801 [hep-ph]} \BibitemShut {NoStop}%
\bibitem [{\citenamefont {Guo}\ \emph {et~al.}(2026)\citenamefont {Guo},
  \citenamefont {Han}, \citenamefont {Low},\ and\ \citenamefont
  {Su}}]{Guo:2026yhz}%
  \BibitemOpen
  \bibfield  {author} {\bibinfo {author} {\bibfnamefont {Y.-C.}\ \bibnamefont
  {Guo}}, \bibinfo {author} {\bibfnamefont {T.}~\bibnamefont {Han}}, \bibinfo
  {author} {\bibfnamefont {M.}~\bibnamefont {Low}}, \ and\ \bibinfo {author}
  {\bibfnamefont {Y.}~\bibnamefont {Su}},\ }\href@noop {} {\  (\bibinfo {year}
  {2026})},\ \Eprint {http://arxiv.org/abs/2602.02719} {arXiv:2602.02719
  [hep-ph]} \BibitemShut {NoStop}%
\bibitem [{\citenamefont {Bernal}\ \emph {et~al.}(2025)\citenamefont {Bernal},
  \citenamefont {Caban},\ and\ \citenamefont
  {Rembieli{\'n}ski}}]{Bernal:2024xhm}%
  \BibitemOpen
  \bibfield  {author} {\bibinfo {author} {\bibfnamefont {A.}~\bibnamefont
  {Bernal}}, \bibinfo {author} {\bibfnamefont {P.}~\bibnamefont {Caban}}, \
  and\ \bibinfo {author} {\bibfnamefont {J.}~\bibnamefont {Rembieli{\'n}ski}},\
  }\href {\doibase 10.1038/s41598-025-07747-3} {\bibfield  {journal} {\bibinfo
  {journal} {Sci. Rep.}\ }\textbf {\bibinfo {volume} {15}},\ \bibinfo {pages}
  {23410} (\bibinfo {year} {2025})},\ \Eprint {http://arxiv.org/abs/2405.16525}
  {arXiv:2405.16525 [hep-ph]} \BibitemShut {NoStop}%
\bibitem [{\citenamefont {Del~Gratta}\ \emph {et~al.}(2025)\citenamefont
  {Del~Gratta}, \citenamefont {Fabbri}, \citenamefont {Lamba}, \citenamefont
  {Maltoni},\ and\ \citenamefont {Pagani}}]{DelGratta:2025qyp}%
  \BibitemOpen
  \bibfield  {author} {\bibinfo {author} {\bibfnamefont {M.}~\bibnamefont
  {Del~Gratta}}, \bibinfo {author} {\bibfnamefont {F.}~\bibnamefont {Fabbri}},
  \bibinfo {author} {\bibfnamefont {P.}~\bibnamefont {Lamba}}, \bibinfo
  {author} {\bibfnamefont {F.}~\bibnamefont {Maltoni}}, \ and\ \bibinfo
  {author} {\bibfnamefont {D.}~\bibnamefont {Pagani}},\ }\href {\doibase
  10.1007/JHEP09(2025)013} {\bibfield  {journal} {\bibinfo  {journal} {JHEP}\
  }\textbf {\bibinfo {volume} {09}},\ \bibinfo {pages} {013} (\bibinfo {year}
  {2025})},\ \Eprint {http://arxiv.org/abs/2504.03841} {arXiv:2504.03841
  [hep-ph]} \BibitemShut {NoStop}%
\bibitem [{\citenamefont {Gon{\c{c}}alves}\ \emph
  {et~al.}(2025{\natexlab{a}})\citenamefont {Gon{\c{c}}alves}, \citenamefont
  {Kaladharan}, \citenamefont {Krauss},\ and\ \citenamefont
  {Navarro}}]{Goncalves:2025mvl}%
  \BibitemOpen
  \bibfield  {author} {\bibinfo {author} {\bibfnamefont {D.}~\bibnamefont
  {Gon{\c{c}}alves}}, \bibinfo {author} {\bibfnamefont {A.}~\bibnamefont
  {Kaladharan}}, \bibinfo {author} {\bibfnamefont {F.}~\bibnamefont {Krauss}},
  \ and\ \bibinfo {author} {\bibfnamefont {A.}~\bibnamefont {Navarro}},\ }\href
  {\doibase 10.1007/JHEP12(2025)122} {\bibfield  {journal} {\bibinfo  {journal}
  {JHEP}\ }\textbf {\bibinfo {volume} {12}},\ \bibinfo {pages} {122} (\bibinfo
  {year} {2025}{\natexlab{a}})},\ \Eprint {http://arxiv.org/abs/2505.12125}
  {arXiv:2505.12125 [hep-ph]} \BibitemShut {NoStop}%
\bibitem [{\citenamefont {Ruzi}\ \emph {et~al.}(2026)\citenamefont {Ruzi},
  \citenamefont {Wu}, \citenamefont {Ding},\ and\ \citenamefont
  {Li}}]{Ruzi:2025jql}%
  \BibitemOpen
  \bibfield  {author} {\bibinfo {author} {\bibfnamefont {A.}~\bibnamefont
  {Ruzi}}, \bibinfo {author} {\bibfnamefont {Y.}~\bibnamefont {Wu}}, \bibinfo
  {author} {\bibfnamefont {R.}~\bibnamefont {Ding}}, \ and\ \bibinfo {author}
  {\bibfnamefont {Q.}~\bibnamefont {Li}},\ }\href {\doibase
  10.1088/1674-1137/ae1374} {\bibfield  {journal} {\bibinfo  {journal} {Chin.
  Phys.}\ }\textbf {\bibinfo {volume} {50}},\ \bibinfo {pages} {023103}
  (\bibinfo {year} {2026})},\ \Eprint {http://arxiv.org/abs/2506.16077}
  {arXiv:2506.16077 [hep-ph]} \BibitemShut {NoStop}%
\bibitem [{\citenamefont {Hong}\ \emph {et~al.}(2025)\citenamefont {Hong},
  \citenamefont {Ping},\ and\ \citenamefont {Song}}]{Hong:2025drg}%
  \BibitemOpen
  \bibfield  {author} {\bibinfo {author} {\bibfnamefont {P.}~\bibnamefont
  {Hong}}, \bibinfo {author} {\bibfnamefont {R.}~\bibnamefont {Ping}}, \ and\
  \bibinfo {author} {\bibfnamefont {W.}~\bibnamefont {Song}},\ }\href@noop {}
  {\  (\bibinfo {year} {2025})},\ \Eprint {http://arxiv.org/abs/2512.22837}
  {arXiv:2512.22837 [hep-ph]} \BibitemShut {NoStop}%
\bibitem [{\citenamefont {Gon{\c{c}}alves}\ \emph
  {et~al.}(2025{\natexlab{b}})\citenamefont {Gon{\c{c}}alves}, \citenamefont
  {Kaladharan},\ and\ \citenamefont {Navarro}}]{Goncalves:2025xer}%
  \BibitemOpen
  \bibfield  {author} {\bibinfo {author} {\bibfnamefont {D.}~\bibnamefont
  {Gon{\c{c}}alves}}, \bibinfo {author} {\bibfnamefont {A.}~\bibnamefont
  {Kaladharan}}, \ and\ \bibinfo {author} {\bibfnamefont {A.}~\bibnamefont
  {Navarro}},\ }\href {\doibase 10.1007/JHEP11(2025)158} {\bibfield  {journal}
  {\bibinfo  {journal} {JHEP}\ }\textbf {\bibinfo {volume} {11}},\ \bibinfo
  {pages} {158} (\bibinfo {year} {2025}{\natexlab{b}})},\ \Eprint
  {http://arxiv.org/abs/2506.19951} {arXiv:2506.19951 [hep-ph]} \BibitemShut
  {NoStop}%
\bibitem [{\citenamefont {Lo~Chiatto}(2025)}]{LoChiatto:2024dmx}%
  \BibitemOpen
  \bibfield  {author} {\bibinfo {author} {\bibfnamefont {P.}~\bibnamefont
  {Lo~Chiatto}},\ }\href {\doibase 10.1103/8gtq-twfc} {\bibfield  {journal}
  {\bibinfo  {journal} {Phys. Rev. D}\ }\textbf {\bibinfo {volume} {112}},\
  \bibinfo {pages} {015017} (\bibinfo {year} {2025})},\ \Eprint
  {http://arxiv.org/abs/2408.04553} {arXiv:2408.04553 [hep-ph]} \BibitemShut
  {NoStop}%
\bibitem [{\citenamefont {Ding}\ \emph {et~al.}(2025)\citenamefont {Ding},
  \citenamefont {Ruzi}, \citenamefont {Qian}, \citenamefont {Levin},
  \citenamefont {Wu},\ and\ \citenamefont {Li}}]{Ding:2025mzj}%
  \BibitemOpen
  \bibfield  {author} {\bibinfo {author} {\bibfnamefont {R.}~\bibnamefont
  {Ding}}, \bibinfo {author} {\bibfnamefont {A.}~\bibnamefont {Ruzi}}, \bibinfo
  {author} {\bibfnamefont {S.}~\bibnamefont {Qian}}, \bibinfo {author}
  {\bibfnamefont {A.}~\bibnamefont {Levin}}, \bibinfo {author} {\bibfnamefont
  {Y.}~\bibnamefont {Wu}}, \ and\ \bibinfo {author} {\bibfnamefont
  {Q.}~\bibnamefont {Li}},\ }\href@noop {} {\  (\bibinfo {year} {2025})},\
  \Eprint {http://arxiv.org/abs/2504.09832} {arXiv:2504.09832 [hep-ph]}
  \BibitemShut {NoStop}%
\bibitem [{\citenamefont {Pei}\ \emph {et~al.}(2025{\natexlab{a}})\citenamefont
  {Pei}, \citenamefont {Hao}, \citenamefont {Wang},\ and\ \citenamefont
  {Li}}]{Pei:2025yvr}%
  \BibitemOpen
  \bibfield  {author} {\bibinfo {author} {\bibfnamefont {J.}~\bibnamefont
  {Pei}}, \bibinfo {author} {\bibfnamefont {X.}~\bibnamefont {Hao}}, \bibinfo
  {author} {\bibfnamefont {X.}~\bibnamefont {Wang}}, \ and\ \bibinfo {author}
  {\bibfnamefont {T.}~\bibnamefont {Li}},\ }\href@noop {} {\  (\bibinfo {year}
  {2025}{\natexlab{a}})},\ \Eprint {http://arxiv.org/abs/2505.09931}
  {arXiv:2505.09931 [hep-ph]} \BibitemShut {NoStop}%
\bibitem [{\citenamefont {Pei}\ \emph {et~al.}(2025{\natexlab{b}})\citenamefont
  {Pei}, \citenamefont {Li}, \citenamefont {Wu}, \citenamefont {Hao},\ and\
  \citenamefont {Wang}}]{Pei:2025ito}%
  \BibitemOpen
  \bibfield  {author} {\bibinfo {author} {\bibfnamefont {J.}~\bibnamefont
  {Pei}}, \bibinfo {author} {\bibfnamefont {T.}~\bibnamefont {Li}}, \bibinfo
  {author} {\bibfnamefont {L.}~\bibnamefont {Wu}}, \bibinfo {author}
  {\bibfnamefont {X.}~\bibnamefont {Hao}}, \ and\ \bibinfo {author}
  {\bibfnamefont {X.}~\bibnamefont {Wang}},\ }\href@noop {} {\  (\bibinfo
  {year} {2025}{\natexlab{b}})},\ \Eprint {http://arxiv.org/abs/2510.08031}
  {arXiv:2510.08031 [hep-ph]} \BibitemShut {NoStop}%
\bibitem [{\citenamefont {Cao}\ \emph {et~al.}(2025)\citenamefont {Cao},
  \citenamefont {Li}, \citenamefont {Wen},\ and\ \citenamefont
  {Yan}}]{Cao:2025qua}%
  \BibitemOpen
  \bibfield  {author} {\bibinfo {author} {\bibfnamefont {Q.-H.}\ \bibnamefont
  {Cao}}, \bibinfo {author} {\bibfnamefont {G.}~\bibnamefont {Li}}, \bibinfo
  {author} {\bibfnamefont {X.-K.}\ \bibnamefont {Wen}}, \ and\ \bibinfo
  {author} {\bibfnamefont {B.}~\bibnamefont {Yan}},\ }\href@noop {} {\
  (\bibinfo {year} {2025})},\ \Eprint {http://arxiv.org/abs/2509.18276}
  {arXiv:2509.18276 [hep-ph]} \BibitemShut {NoStop}%
\bibitem [{\citenamefont {Cheng}\ \emph
  {et~al.}(2025{\natexlab{c}})\citenamefont {Cheng}, \citenamefont {Han},
  \citenamefont {Low},\ and\ \citenamefont {Wu}}]{Cheng:2025zcf}%
  \BibitemOpen
  \bibfield  {author} {\bibinfo {author} {\bibfnamefont {K.}~\bibnamefont
  {Cheng}}, \bibinfo {author} {\bibfnamefont {T.}~\bibnamefont {Han}}, \bibinfo
  {author} {\bibfnamefont {M.}~\bibnamefont {Low}}, \ and\ \bibinfo {author}
  {\bibfnamefont {T.~A.}\ \bibnamefont {Wu}},\ }\href@noop {} {\  (\bibinfo
  {year} {2025}{\natexlab{c}})},\ \Eprint {http://arxiv.org/abs/2507.12513}
  {arXiv:2507.12513 [hep-ph]} \BibitemShut {NoStop}%
\bibitem [{\citenamefont {Shi}\ and\ \citenamefont {Yang}(2018)}]{Shi:2016bvo}%
  \BibitemOpen
  \bibfield  {author} {\bibinfo {author} {\bibfnamefont {Y.}~\bibnamefont
  {Shi}}\ and\ \bibinfo {author} {\bibfnamefont {J.}~\bibnamefont {Yang}},\
  }\href {\doibase 10.1103/PhysRevD.98.075019} {\bibfield  {journal} {\bibinfo
  {journal} {Phys. Rev. D}\ }\textbf {\bibinfo {volume} {98}},\ \bibinfo
  {pages} {075019} (\bibinfo {year} {2018})},\ \Eprint
  {http://arxiv.org/abs/1612.07628} {arXiv:1612.07628 [hep-ph]} \BibitemShut
  {NoStop}%
\bibitem [{\citenamefont {Shi}\ and\ \citenamefont
  {Yang}(2020{\natexlab{a}})}]{Shi:2019mlf}%
  \BibitemOpen
  \bibfield  {author} {\bibinfo {author} {\bibfnamefont {Y.}~\bibnamefont
  {Shi}}\ and\ \bibinfo {author} {\bibfnamefont {J.-C.}\ \bibnamefont {Yang}},\
  }\href {\doibase 10.1140/epjc/s10052-020-08430-9} {\bibfield  {journal}
  {\bibinfo  {journal} {Eur. Phys. J. C}\ }\textbf {\bibinfo {volume} {80}},\
  \bibinfo {pages} {861} (\bibinfo {year} {2020}{\natexlab{a}})},\ \Eprint
  {http://arxiv.org/abs/1909.10626} {arXiv:1909.10626 [hep-ph]} \BibitemShut
  {NoStop}%
\bibitem [{\citenamefont {Shi}\ and\ \citenamefont
  {Yang}(2020{\natexlab{b}})}]{Shi:2019kjf}%
  \BibitemOpen
  \bibfield  {author} {\bibinfo {author} {\bibfnamefont {Y.}~\bibnamefont
  {Shi}}\ and\ \bibinfo {author} {\bibfnamefont {J.-C.}\ \bibnamefont {Yang}},\
  }\href {\doibase 10.1140/epjc/s10052-020-7684-5} {\bibfield  {journal}
  {\bibinfo  {journal} {Eur. Phys. J. C}\ }\textbf {\bibinfo {volume} {80}},\
  \bibinfo {pages} {116} (\bibinfo {year} {2020}{\natexlab{b}})},\ \Eprint
  {http://arxiv.org/abs/1912.04111} {arXiv:1912.04111 [hep-ph]} \BibitemShut
  {NoStop}%
\bibitem [{\citenamefont {Pei}\ \emph {et~al.}(2026{\natexlab{a}})\citenamefont
  {Pei}, \citenamefont {Wu}, \citenamefont {Wang}, \citenamefont {Hao},\ and\
  \citenamefont {Li}}]{Pei:2026rlh}%
  \BibitemOpen
  \bibfield  {author} {\bibinfo {author} {\bibfnamefont {J.}~\bibnamefont
  {Pei}}, \bibinfo {author} {\bibfnamefont {L.}~\bibnamefont {Wu}}, \bibinfo
  {author} {\bibfnamefont {D.}~\bibnamefont {Wang}}, \bibinfo {author}
  {\bibfnamefont {X.}~\bibnamefont {Hao}}, \ and\ \bibinfo {author}
  {\bibfnamefont {T.}~\bibnamefont {Li}},\ }\href@noop {} {\  (\bibinfo {year}
  {2026}{\natexlab{a}})},\ \Eprint {http://arxiv.org/abs/2601.15748}
  {arXiv:2601.15748 [hep-ph]} \BibitemShut {NoStop}%
\bibitem [{\citenamefont {Pei}\ \emph {et~al.}(2026{\natexlab{b}})\citenamefont
  {Pei}, \citenamefont {Wu}, \citenamefont {Li},\ and\ \citenamefont
  {Hao}}]{Pei:2026wfu}%
  \BibitemOpen
  \bibfield  {author} {\bibinfo {author} {\bibfnamefont {J.}~\bibnamefont
  {Pei}}, \bibinfo {author} {\bibfnamefont {L.}~\bibnamefont {Wu}}, \bibinfo
  {author} {\bibfnamefont {T.}~\bibnamefont {Li}}, \ and\ \bibinfo {author}
  {\bibfnamefont {X.}~\bibnamefont {Hao}},\ }\href@noop {} {\  (\bibinfo {year}
  {2026}{\natexlab{b}})},\ \Eprint {http://arxiv.org/abs/2601.15747}
  {arXiv:2601.15747 [hep-ph]} \BibitemShut {NoStop}%
\bibitem [{\citenamefont {Pei}\ and\ \citenamefont {Wu}(2026)}]{Pei:2026khg}%
  \BibitemOpen
  \bibfield  {author} {\bibinfo {author} {\bibfnamefont {J.}~\bibnamefont
  {Pei}}\ and\ \bibinfo {author} {\bibfnamefont {L.}~\bibnamefont {Wu}},\
  }\href@noop {} {\  (\bibinfo {year} {2026})},\ \Eprint
  {http://arxiv.org/abs/2602.08541} {arXiv:2602.08541 [hep-ph]} \BibitemShut
  {NoStop}%
\bibitem [{\citenamefont {Antozzi}\ \emph {et~al.}(2026)\citenamefont
  {Antozzi}, \citenamefont {Chalbaud}, \citenamefont {D{\'e}liot},
  \citenamefont {Fabbri}, \citenamefont {Fiolhais}, \citenamefont {Fuks},
  \citenamefont {Onofre}, \citenamefont {White},\ and\ \citenamefont
  {Zhu}}]{Antozzi:2026vdi}%
  \BibitemOpen
  \bibfield  {author} {\bibinfo {author} {\bibfnamefont {L.}~\bibnamefont
  {Antozzi}}, \bibinfo {author} {\bibfnamefont {E.}~\bibnamefont {Chalbaud}},
  \bibinfo {author} {\bibfnamefont {F.}~\bibnamefont {D{\'e}liot}}, \bibinfo
  {author} {\bibfnamefont {F.}~\bibnamefont {Fabbri}}, \bibinfo {author}
  {\bibfnamefont {M.~C.~N.}\ \bibnamefont {Fiolhais}}, \bibinfo {author}
  {\bibfnamefont {B.}~\bibnamefont {Fuks}}, \bibinfo {author} {\bibfnamefont
  {A.}~\bibnamefont {Onofre}}, \bibinfo {author} {\bibfnamefont
  {M.}~\bibnamefont {White}}, \ and\ \bibinfo {author} {\bibfnamefont
  {P.}~\bibnamefont {Zhu}},\ }\href@noop {} {\  (\bibinfo {year} {2026})},\
  \Eprint {http://arxiv.org/abs/2602.23426} {arXiv:2602.23426 [hep-ph]}
  \BibitemShut {NoStop}%
\bibitem [{\citenamefont {Barr}\ \emph {et~al.}(2024)\citenamefont {Barr},
  \citenamefont {Fabbrichesi}, \citenamefont {Floreanini}, \citenamefont
  {Gabrielli},\ and\ \citenamefont {Marzola}}]{Barr:2024djo}%
  \BibitemOpen
  \bibfield  {author} {\bibinfo {author} {\bibfnamefont {A.~J.}\ \bibnamefont
  {Barr}}, \bibinfo {author} {\bibfnamefont {M.}~\bibnamefont {Fabbrichesi}},
  \bibinfo {author} {\bibfnamefont {R.}~\bibnamefont {Floreanini}}, \bibinfo
  {author} {\bibfnamefont {E.}~\bibnamefont {Gabrielli}}, \ and\ \bibinfo
  {author} {\bibfnamefont {L.}~\bibnamefont {Marzola}},\ }\href {\doibase
  10.1016/j.ppnp.2024.104134} {\bibfield  {journal} {\bibinfo  {journal} {Prog.
  Part. Nucl. Phys.}\ }\textbf {\bibinfo {volume} {139}},\ \bibinfo {pages}
  {104134} (\bibinfo {year} {2024})},\ \Eprint
  {http://arxiv.org/abs/2402.07972} {arXiv:2402.07972 [hep-ph]} \BibitemShut
  {NoStop}%
\bibitem [{\citenamefont {Go}(2004)}]{Go:2003tx}%
  \BibitemOpen
  \bibfield  {author} {\bibinfo {author} {\bibfnamefont {A.}~\bibnamefont
  {Go}},\ }\href {\doibase 10.1080/09500340408233614} {\bibfield  {journal}
  {\bibinfo  {journal} {J. Mod. Opt.}\ }\textbf {\bibinfo {volume} {51}},\
  \bibinfo {pages} {991} (\bibinfo {year} {2004})},\ \Eprint
  {http://arxiv.org/abs/quant-ph/0310192} {arXiv:quant-ph/0310192} \BibitemShut
  {NoStop}%
\bibitem [{\citenamefont {Achasov}\ \emph {et~al.}(2024)\citenamefont {Achasov}
  \emph {et~al.}}]{Achasov:2023gey}%
  \BibitemOpen
  \bibfield  {author} {\bibinfo {author} {\bibfnamefont {M.}~\bibnamefont
  {Achasov}} \emph {et~al.},\ }\href {\doibase 10.1007/s11467-023-1333-z}
  {\bibfield  {journal} {\bibinfo  {journal} {Front. Phys. (Beijing)}\ }\textbf
  {\bibinfo {volume} {19}},\ \bibinfo {pages} {14701} (\bibinfo {year}
  {2024})},\ \Eprint {http://arxiv.org/abs/2303.15790} {arXiv:2303.15790
  [hep-ex]} \BibitemShut {NoStop}%
\bibitem [{\citenamefont {Ai}\ \emph {et~al.}(2025)\citenamefont {Ai} \emph
  {et~al.}}]{Ai:2025xop}%
  \BibitemOpen
  \bibfield  {author} {\bibinfo {author} {\bibfnamefont {X.-C.}\ \bibnamefont
  {Ai}} \emph {et~al.},\ }\href {\doibase 10.1007/s41365-025-01833-x}
  {\bibfield  {journal} {\bibinfo  {journal} {Nucl. Sci. Tech.}\ }\textbf
  {\bibinfo {volume} {36}},\ \bibinfo {pages} {242} (\bibinfo {year} {2025})},\
  \Eprint {http://arxiv.org/abs/2509.11522} {arXiv:2509.11522 [physics.acc-ph]}
  \BibitemShut {NoStop}%
\bibitem [{\citenamefont {Ablikim}\ \emph {et~al.}(2010)\citenamefont {Ablikim}
  \emph {et~al.}}]{BESIII:2009fln}%
  \BibitemOpen
  \bibfield  {author} {\bibinfo {author} {\bibfnamefont {M.}~\bibnamefont
  {Ablikim}} \emph {et~al.} (\bibinfo {collaboration} {BESIII}),\ }\href
  {\doibase 10.1016/j.nima.2009.12.050} {\bibfield  {journal} {\bibinfo
  {journal} {Nucl. Instrum. Meth. A}\ }\textbf {\bibinfo {volume} {614}},\
  \bibinfo {pages} {345} (\bibinfo {year} {2010})},\ \Eprint
  {http://arxiv.org/abs/0911.4960} {arXiv:0911.4960 [physics.ins-det]}
  \BibitemShut {NoStop}%
\bibitem [{\citenamefont {Privitera}(1992)}]{Privitera:1991nz}%
  \BibitemOpen
  \bibfield  {author} {\bibinfo {author} {\bibfnamefont {P.}~\bibnamefont
  {Privitera}},\ }\href {\doibase 10.1016/0370-2693(92)90872-2} {\bibfield
  {journal} {\bibinfo  {journal} {Phys. Lett. B}\ }\textbf {\bibinfo {volume}
  {275}},\ \bibinfo {pages} {172} (\bibinfo {year} {1992})}\BibitemShut
  {NoStop}%
\bibitem [{\citenamefont {Dreiner}(1992)}]{Dreiner:1992gt}%
  \BibitemOpen
  \bibfield  {author} {\bibinfo {author} {\bibfnamefont {H.~K.}\ \bibnamefont
  {Dreiner}},\ }in\ \href@noop {} {\emph {\bibinfo {booktitle} {{2nd Workshop
  on Tau Lepton Physics}}}}\ (\bibinfo {year} {1992})\ \Eprint
  {http://arxiv.org/abs/hep-ph/9211203} {arXiv:hep-ph/9211203} \BibitemShut
  {NoStop}%
\bibitem [{\citenamefont {Zhang}\ \emph {et~al.}(2025)\citenamefont {Zhang},
  \citenamefont {Zhou}, \citenamefont {Liu}, \citenamefont {Wu}, \citenamefont
  {Li}, \citenamefont {Han}, \citenamefont {Hsu},\ and\ \citenamefont
  {Low}}]{Zhang:2025mmm}%
  \BibitemOpen
  \bibfield  {author} {\bibinfo {author} {\bibfnamefont {Y.}~\bibnamefont
  {Zhang}}, \bibinfo {author} {\bibfnamefont {B.-H.}\ \bibnamefont {Zhou}},
  \bibinfo {author} {\bibfnamefont {Q.-B.}\ \bibnamefont {Liu}}, \bibinfo
  {author} {\bibfnamefont {T.~A.}\ \bibnamefont {Wu}}, \bibinfo {author}
  {\bibfnamefont {S.}~\bibnamefont {Li}}, \bibinfo {author} {\bibfnamefont
  {T.}~\bibnamefont {Han}}, \bibinfo {author} {\bibfnamefont {S.-C.}\
  \bibnamefont {Hsu}}, \ and\ \bibinfo {author} {\bibfnamefont
  {M.}~\bibnamefont {Low}},\ }\href@noop {} {\  (\bibinfo {year} {2025})},\
  \Eprint {http://arxiv.org/abs/2504.01496} {arXiv:2504.01496 [hep-ph]}
  \BibitemShut {NoStop}%
\bibitem [{\citenamefont {Lu}\ \emph {et~al.}(2025)\citenamefont {Lu},
  \citenamefont {Si}, \citenamefont {Zhang},\ and\ \citenamefont
  {Zhang}}]{Lu:2025hwy}%
  \BibitemOpen
  \bibfield  {author} {\bibinfo {author} {\bibfnamefont {P.-C.}\ \bibnamefont
  {Lu}}, \bibinfo {author} {\bibfnamefont {Z.-G.}\ \bibnamefont {Si}}, \bibinfo
  {author} {\bibfnamefont {H.}~\bibnamefont {Zhang}}, \ and\ \bibinfo {author}
  {\bibfnamefont {X.-Y.}\ \bibnamefont {Zhang}},\ }\href@noop {} {\  (\bibinfo
  {year} {2025})},\ \Eprint {http://arxiv.org/abs/2511.18935} {arXiv:2511.18935
  [hep-ph]} \BibitemShut {NoStop}%
\bibitem [{\citenamefont {Fabbrichesi}\ and\ \citenamefont
  {Marzola}(2024)}]{Fabbrichesi:2024wcd}%
  \BibitemOpen
  \bibfield  {author} {\bibinfo {author} {\bibfnamefont {M.}~\bibnamefont
  {Fabbrichesi}}\ and\ \bibinfo {author} {\bibfnamefont {L.}~\bibnamefont
  {Marzola}},\ }\href {\doibase 10.1103/PhysRevD.110.076004} {\bibfield
  {journal} {\bibinfo  {journal} {Phys. Rev. D}\ }\textbf {\bibinfo {volume}
  {110}},\ \bibinfo {pages} {076004} (\bibinfo {year} {2024})},\ \Eprint
  {http://arxiv.org/abs/2405.09201} {arXiv:2405.09201 [hep-ph]} \BibitemShut
  {NoStop}%
\bibitem [{\citenamefont {Bechtle}\ \emph {et~al.}(2025)\citenamefont
  {Bechtle}, \citenamefont {Breuning}, \citenamefont {Dreiner},\ and\
  \citenamefont {Duhr}}]{Bechtle:2025ugc}%
  \BibitemOpen
  \bibfield  {author} {\bibinfo {author} {\bibfnamefont {P.}~\bibnamefont
  {Bechtle}}, \bibinfo {author} {\bibfnamefont {C.}~\bibnamefont {Breuning}},
  \bibinfo {author} {\bibfnamefont {H.~K.}\ \bibnamefont {Dreiner}}, \ and\
  \bibinfo {author} {\bibfnamefont {C.}~\bibnamefont {Duhr}},\ }\href@noop {}
  {\  (\bibinfo {year} {2025})},\ \Eprint {http://arxiv.org/abs/2507.15947}
  {arXiv:2507.15947 [hep-ph]} \BibitemShut {NoStop}%
\bibitem [{\citenamefont {Abel}\ \emph {et~al.}(2025)\citenamefont {Abel},
  \citenamefont {Dreiner}, \citenamefont {Sengupta},\ and\ \citenamefont
  {Ubaldi}}]{Abel:2025skj}%
  \BibitemOpen
  \bibfield  {author} {\bibinfo {author} {\bibfnamefont {S.~A.}\ \bibnamefont
  {Abel}}, \bibinfo {author} {\bibfnamefont {H.~K.}\ \bibnamefont {Dreiner}},
  \bibinfo {author} {\bibfnamefont {R.}~\bibnamefont {Sengupta}}, \ and\
  \bibinfo {author} {\bibfnamefont {L.}~\bibnamefont {Ubaldi}},\ }\href@noop {}
  {\  (\bibinfo {year} {2025})},\ \Eprint {http://arxiv.org/abs/2507.15949}
  {arXiv:2507.15949 [hep-ph]} \BibitemShut {NoStop}%
\bibitem [{\citenamefont {Li}\ \emph {et~al.}(2024)\citenamefont {Li},
  \citenamefont {Shen},\ and\ \citenamefont {Yang}}]{Li:2024luk}%
  \BibitemOpen
  \bibfield  {author} {\bibinfo {author} {\bibfnamefont {S.}~\bibnamefont
  {Li}}, \bibinfo {author} {\bibfnamefont {W.}~\bibnamefont {Shen}}, \ and\
  \bibinfo {author} {\bibfnamefont {J.~M.}\ \bibnamefont {Yang}},\ }\href
  {\doibase 10.1140/epjc/s10052-024-13584-x} {\bibfield  {journal} {\bibinfo
  {journal} {Eur. Phys. J. C}\ }\textbf {\bibinfo {volume} {84}},\ \bibinfo
  {pages} {1195} (\bibinfo {year} {2024})},\ \Eprint
  {http://arxiv.org/abs/2401.01162} {arXiv:2401.01162 [hep-th]} \BibitemShut
  {NoStop}%
\bibitem [{\citenamefont {Low}(2025)}]{Low:2025aqq}%
  \BibitemOpen
  \bibfield  {author} {\bibinfo {author} {\bibfnamefont {M.}~\bibnamefont
  {Low}},\ }\href {\doibase 10.1103/15c3-mg5l} {\bibfield  {journal} {\bibinfo
  {journal} {Phys. Rev. D}\ }\textbf {\bibinfo {volume} {112}},\ \bibinfo
  {pages} {096008} (\bibinfo {year} {2025})},\ \Eprint
  {http://arxiv.org/abs/2508.10979} {arXiv:2508.10979 [hep-ph]} \BibitemShut
  {NoStop}%
\bibitem [{\citenamefont {Fano}(1983)}]{Fano:1983zz}%
  \BibitemOpen
  \bibfield  {author} {\bibinfo {author} {\bibfnamefont {U.}~\bibnamefont
  {Fano}},\ }\href {\doibase 10.1103/RevModPhys.55.855} {\bibfield  {journal}
  {\bibinfo  {journal} {Rev. Mod. Phys.}\ }\textbf {\bibinfo {volume} {55}},\
  \bibinfo {pages} {855} (\bibinfo {year} {1983})}\BibitemShut {NoStop}%
\bibitem [{\citenamefont {Jeans}(2016)}]{Jeans:2015vaa}%
  \BibitemOpen
  \bibfield  {author} {\bibinfo {author} {\bibfnamefont {D.}~\bibnamefont
  {Jeans}},\ }\href {\doibase 10.1016/j.nima.2015.11.030} {\bibfield  {journal}
  {\bibinfo  {journal} {Nucl. Instrum. Meth. A}\ }\textbf {\bibinfo {volume}
  {810}},\ \bibinfo {pages} {51} (\bibinfo {year} {2016})},\ \Eprint
  {http://arxiv.org/abs/1507.01700} {arXiv:1507.01700 [hep-ex]} \BibitemShut
  {NoStop}%
\bibitem [{\citenamefont {Bodrov}(2024)}]{Bodrov:2024wrw}%
  \BibitemOpen
  \bibfield  {author} {\bibinfo {author} {\bibfnamefont {D.}~\bibnamefont
  {Bodrov}},\ }\href {\doibase 10.1142/S0217751X24420065} {\bibfield  {journal}
  {\bibinfo  {journal} {Int. J. Mod. Phys. A}\ }\textbf {\bibinfo {volume}
  {39}},\ \bibinfo {pages} {2442006} (\bibinfo {year} {2024})},\ \Eprint
  {http://arxiv.org/abs/2405.16512} {arXiv:2405.16512 [hep-ex]} \BibitemShut
  {NoStop}%
\bibitem [{\citenamefont {Kuhn}(1993)}]{Kuhn:1993ra}%
  \BibitemOpen
  \bibfield  {author} {\bibinfo {author} {\bibfnamefont {J.~H.}\ \bibnamefont
  {Kuhn}},\ }\href {\doibase 10.1016/0370-2693(93)90019-E} {\bibfield
  {journal} {\bibinfo  {journal} {Phys. Lett. B}\ }\textbf {\bibinfo {volume}
  {313}},\ \bibinfo {pages} {458} (\bibinfo {year} {1993})},\ \Eprint
  {http://arxiv.org/abs/hep-ph/9307269} {arXiv:hep-ph/9307269} \BibitemShut
  {NoStop}%
\bibitem [{\citenamefont {Sun}\ \emph {et~al.}(2025)\citenamefont {Sun},
  \citenamefont {Wu},\ and\ \citenamefont {Zhou}}]{Sun:2024vcd}%
  \BibitemOpen
  \bibfield  {author} {\bibinfo {author} {\bibfnamefont {X.}~\bibnamefont
  {Sun}}, \bibinfo {author} {\bibfnamefont {Y.}~\bibnamefont {Wu}}, \ and\
  \bibinfo {author} {\bibfnamefont {X.}~\bibnamefont {Zhou}},\ }\href {\doibase
  10.1088/1674-1137/adf6e0} {\bibfield  {journal} {\bibinfo  {journal} {Chin.
  Phys.}\ }\textbf {\bibinfo {volume} {49}},\ \bibinfo {pages} {113001}
  (\bibinfo {year} {2025})},\ \Eprint {http://arxiv.org/abs/2411.19469}
  {arXiv:2411.19469 [hep-ex]} \BibitemShut {NoStop}%
\bibitem [{\citenamefont {Navas}\ \emph {et~al.}(2024)\citenamefont {Navas}
  \emph {et~al.}}]{ParticleDataGroup:2024cfk}%
  \BibitemOpen
  \bibfield  {author} {\bibinfo {author} {\bibfnamefont {S.}~\bibnamefont
  {Navas}} \emph {et~al.} (\bibinfo {collaboration} {Particle Data Group}),\
  }\href {\doibase 10.1103/PhysRevD.110.030001} {\bibfield  {journal} {\bibinfo
   {journal} {Phys. Rev. D}\ }\textbf {\bibinfo {volume} {110}},\ \bibinfo
  {pages} {030001} (\bibinfo {year} {2024})}\BibitemShut {NoStop}%
\bibitem [{\citenamefont {Adachi}\ \emph {et~al.}(2023)\citenamefont {Adachi}
  \emph {et~al.}}]{Belle-II:2023izd}%
  \BibitemOpen
  \bibfield  {author} {\bibinfo {author} {\bibfnamefont {I.}~\bibnamefont
  {Adachi}} \emph {et~al.} (\bibinfo {collaboration} {Belle-II}),\ }\href
  {\doibase 10.1103/PhysRevD.108.032006} {\bibfield  {journal} {\bibinfo
  {journal} {Phys. Rev. D}\ }\textbf {\bibinfo {volume} {108}},\ \bibinfo
  {pages} {032006} (\bibinfo {year} {2023})},\ \Eprint
  {http://arxiv.org/abs/2305.19116} {arXiv:2305.19116 [hep-ex]} \BibitemShut
  {NoStop}%
\end{thebibliography}%

\end{document}